\documentclass[acmsmall]{acmart}
\AtBeginDocument{%
  \providecommand\BibTeX{{%
    \normalfont B\kern-0.5em{\scshape i\kern-0.25em b}\kern-0.8em\TeX}}}

\setcopyright{acmlicensed}
\copyrightyear{2025}
\acmYear{2025}
\acmDOI{XXXXXXX.XXXXXXX}

\usepackage{multirow}
\usepackage{tabularx}
\usepackage{rotating}
\usepackage{xcolor}
\usepackage{amsthm}

\usepackage{amssymb}
\usepackage{colortbl}
\usepackage{pifont}

\usepackage[normalem]{ulem}

\useunder{\uline}{\ul}{}

\newcommand{\cmark}{\textcolor{green}{\textbf{\ding{51}}}}
\newcommand{\xmark}{\textcolor{red}{\textbf{\ding{55}}}}
\acmJournal{CSUR}
\acmVolume{9}
\acmNumber{9}
\acmArticle{35}
\acmMonth{9}

\begin{document}

%%
%% The ''title'' command has an optional parameter,
%% allowing the author to define a ''short title'' to be used in page headers.
\title{AI Reasoning for Wireless Communications and Networking: A Survey and Perspectives}

\authorsaddresses{Authors' Contact Information: 
Haoxiang Luo, lhx991115@163.com, University of Electronic Science and Technology of China (UESTC), Chengdu, China;
Yu Yan, 17828358616@163.com, UESTC, Chengdu, China;
Yanhui Bian, yhbian@std.uestc.edu.cn, UESTC, Chengdu, China;
Wenjiao Feng, wenjiaofeng@std.uestc.edu.cn, UESTC, Chengdu, China;
Ruichen Zhang, ruichen.zhang@ntu.edu.sg, Nanyang Technological University (NTU), Singapore;
Yinqiu Liu, yinqiu001@e.ntu.edu.sg, NTU, Singapore;
Jiacheng Wang, jiacheng.wang@ntu.edu.sg, NTU, Singapore;
Gang Sun, gangsun@uestc.edu.cn, UESTC, Chengdu, China;
Dusit Niyato, dniyato@ntu.edu.sg, NTU, Singapore;
Hongfang Yu, yuhf@uestc.edu.cn, UESTC, Chengdu, China;
Abbas Jamalipour, a.jamalipour@ieee.org, The University of Sydney, Sydney, Australia;
Shiwen Mao, smao@ieee.org, Auburn University, Auburn, USA.
}

\thanks{Haoxiang Luo, Yu Yan, and Yanhui Bian contributed equally to this work. The corresponding author: Gang Sun.}

\settopmatter{printacmref=false}
\setcopyright{none}
\renewcommand\footnotetextcopyrightpermission[1]{}
% \pagestyle{plain}

%%
%% The ''author'' command and its associated commands are used to define
% ''authornote'' and ''authornotemark'' commands
% used to denote shared contribution to the research.
% \affiliation{%

\author{Haoxiang Luo}
\affiliation{%
  \institution{University of Electronic Science and Technology of China}
  \country{China}
}

\author{Yu Yan}
\affiliation{%
  \institution{University of Electronic Science and Technology of China}
  \country{China}
}

\author{Yanhui Bian}
\affiliation{%
  \institution{University of Electronic Science and Technology of China}
  \country{China}
}

\author{Wenjiao Feng}
\affiliation{%
  \institution{University of Electronic Science and Technology of China}
  \country{China}
}

\author{Ruichen Zhang}
\affiliation{%
  \institution{Nanyang Technological University}
  \country{Singapore}
}

\author{Yinqiu Liu}
\affiliation{%  
  \institution{Nanyang Technological University}
  \country{Singapore}
}

\author{Jiacheng Wang}
\affiliation{%  
  \institution{Nanyang Technological University}
  \country{Singapore}
}

\author{Gang Sun}
\affiliation{%
  \institution{University of Electronic Science and Technology of China}
  \country{China}
}

\author{Dusit Niyato}
\affiliation{%
  \institution{Nanyang Technological University}
  \country{Singapore}
}

\author{Hongfang Yu}
\affiliation{%
  \institution{University of Electronic Science and Technology of China}
  \country{China}
}

\author{Abbas Jamalipour}
\affiliation{%
  \institution{The University of Sydney}
  \country{Australia}
}

\author{Shiwen Mao}
\affiliation{%
  \institution{Auburn University}
  \country{USA}
}

\renewcommand{\shortauthors}{H. Luo et al.}

\begin{abstract}
Artificial Intelligence (AI) techniques play a pivotal role in optimizing wireless communication networks. However, traditional deep learning approaches often act as closed boxes, lacking the structured reasoning abilities needed to tackle complex, multi-step decision problems. This survey provides a comprehensive review and outlook of reasoning-enabled AI in wireless communication networks, with a focus on Large Language Models (LLMs) and other advanced reasoning paradigms. In particular, LLM-based agents can combine reasoning with long-term planning, memory, tool utilization, and autonomous cross-layer control to dynamically optimize network operations with minimal human intervention. We begin by outlining the evolution of intelligent wireless networking and the limitations of conventional AI methods. We then introduce emerging AI reasoning techniques. Furthermore, we establish a classification system applicable to wireless network tasks. We also present a layer-by-layer examination for AI reasoning, covering the physical, data link, network, transport, and application layers. For each part, we identify key challenges and illustrate how AI reasoning methods can improve AI-based wireless communication performance. Finally, we discuss key research directions for AI reasoning toward future wireless communication networks. By combining insights from both communications and AI, this survey aims to chart a path for integrating reasoning techniques into the next-generation wireless networks.
\end{abstract}





\maketitle

\section{Introduction}
\subsection{Background}
Wireless communications and networking are entering an era of unprecedented complexity and intelligence requirements. Next-generation networks (5G-Advanced and 6G) must support a massive scale of devices, diverse service requirements, and dynamic environments \cite{cui2025overview}. To meet these demands, intelligent wireless networking has emerged as a key technology, leveraging AI and Machine Learning (ML) for tasks ranging from physical-layer signal processing to network management \cite{sun2025comprehensive}. Over the past decades, numerous successes have been reported in applying deep learning to wireless problems, for example, neural networks for channel decoding, traffic prediction, resource allocation, and routing \cite{mao2018deep}. These data-driven methods often outperform traditional handcrafted algorithms under specific conditions. Nonetheless, they also face fundamental limitations when deployed in complex, real-world network scenarios. 

Conventional deep learning models in wireless networks typically function as closed boxes that map inputs to outputs without explicit reasoning or transparency. While being able to learn patterns from training data, they struggle with out-of-distribution generalization and adapting to conditions not seen during training. However, wireless networks are highly variable \cite{luo2023performance}, \cite{luo2024escm}. Channel conditions, user mobility, traffic loads, and network topology can change in unforeseen ways \cite{luo2025convergence}, \cite{luo2024symbiotic}. A fixed neural model may fail to handle new interference scenarios or unseen network topologies, leading to performance degradation. Moreover, deep learning models lack interpretability \cite{zhang2021survey}. Network operators struggle to understand or trust decisions made by AI, which is particularly problematic in mission-critical settings \cite{wang2025chain}, \cite{wang2024generative}. For instance, if a deep learning model adjusts power levels or routing paths, operators need insight into why those actions are chosen. Additionally, traditional AI often requires large labeled datasets for training \cite{roh2019survey}, which are scarce in specialized domains such as wireless. For example, it is impractical to collect exhaustive examples of every possible network condition. These shortcomings motivate the incorporation of reasoning capabilities into AI for wireless, to enable step-by-step logical decision making, better use of prior knowledge, and improved explainability \cite{thomas2024causal}. A typical example is that Cisco Systems is about to launch the world's first intelligent security model with adaptive reasoning capabilities, to address the increasingly complex cybersecurity issues\footnote{https://cybernews.com/ai-news/cisco-first-ever-ai-powered-security-reasoning-model/}.

Recent advances in AI, particularly in LLMs, offer new opportunities to overcome the above limitations \cite{luo2024bc4llm}. LLMs such as GPT-4 have demonstrated an ability to perform complex reasoning by breaking problems into intermediate steps, recalling factual knowledge, and planning actions \cite{tang2024science}, \cite{luo2025toward}. While developed mainly for natural language tasks, such capabilities are now being explored in the context of wireless networks \cite{xu2024large}. The key idea is to imbue AI agents with reasoning strategies, so that they can handle the logical and combinatorial problems that pure end-to-end learning might miss. For example, an LLM-based network controller could parse a high-level goal into low-level configuration steps, namely Chain of Thought (CoT) \cite{gao2025langcoop} or consult an external knowledge base of protocols and past outcomes, that is, Retrieval-Augmented Generation (RAG) \cite{zhang2024interactive}, before making a decision. 
By doing so, the AI’s decisions become more transparent and grounded in known principles, reducing failures due to hallucinations or logical errors \cite{liu2024hallucination}. Furthermore, reasoning allows integration of domain knowledge, such as symbolic rules and analytical models, which can be combined with learning for better reliability \cite{chudasama2025towards}. These developments mark a shift from treating wireless AI as purely a statistical function approximation, towards cognitive AI that can reason, explain, and incorporate knowledge.

Several recent research studies emphasize the importance of reasoning-aware AI in wireless communications. For instance, Wang et al. \cite{wang2025generative} demonstrated that applying CoT prompting, a technique that guides an LLM to generate explicit intermediate reasoning steps, significantly improves an LLM’s ability to handle multi-step tasks in wireless scenarios, such as interference management and beamforming optimization. 
By making the model’s thought process explicit, CoT prompting not only boosted performance but also enhanced interpretability for network operators. Likewise, Zhou et al. \cite{zhou2025large} showed that LLM reasoning methods can enable network optimization and prediction tasks with minimal fine-tuning. It is attractive given the limited computing resources in many network devices. These early successes highlight a trend: LLM-empowered solutions that use reasoning are emerging as a promising approach to tackle wireless networking challenges.

\begin{table*}[tpb]
\centering
\caption{Summary of related surveys}
\vspace{-0.3cm}
\label{tab1}
\tiny
\renewcommand{\arraystretch}{1.4}
% 第一列保留原始的m{1.6cm}格式和双竖线分隔，其他列采用新格式
\begin{tabular}{m{1.5cm}||c|p{6.5cm}|c|c|c}
\hline
\hline
\rowcolor{gray!15} % 添加浅灰色背景
\textbf{Scope} & \textbf{Ref.} & \textbf{Overview} & \textbf{LLM} & \textbf{Reasoning} & \textbf{Wireless scenarios} \\
\hline
\multirow{14}*{AI for Wireless} & \cite{cui2025overview} & A comprehensive overview on AI for network optimization, network for AI enablement, and AI-as-a-service in 6G & \cmark & \xmark & \cmark \\
\cline{2-6}
~ & \cite{zhou2025large1} & An overview about the prompt engineering techniques for adapting LLMs to wireless networks & \cmark & \xmark & \cmark \\
\cline{2-6}
~ & \cite{nguyen2022transfer} & A survey on transfer learning techniques for diverse wireless tasks, such as spectrum management, signal recognition, security, caching, etc. & \xmark & \xmark & \cmark \\
\cline{2-6}
~ & \cite{golec2025llm} & A comprehensive systematic review and classification study on advanced persistent threats detection assisted by LLM in 6G networks & \cmark & \xmark & \cmark \\
\cline{2-6}
~ & \cite{chen2024big} & An overview about foundation models for 6G wireless networks and advocating the design of wireless big AI models & \cmark & \xmark & \cmark \\
\cline{2-6}
~ & \cite{zhou2025large} & A survey introducing the LLM applications in telecom networks (6G and beyond) and identifying future challenges & \cmark & Partially & \cmark \\
\cline{2-6}
~ & \cite{long2025survey} & A survey on LLM development and capabilities for intelligent network operations and LLM-driven performance optimization & \cmark & \xmark & \cmark \\
\cline{2-6}
~ & \cite{boateng2025survey} & A survey on LLM for network and service management across multiple domains, such as mobile, vehicular, cloud/edge & \cmark & \xmark & \cmark \\
\cline{2-6}
~ & \cite{zhu2025wireless} & A comprehensive review clarifying the fundamental principles, diverse applications, and key challenges of wireless large AI models for 6G & \cmark & \xmark & \cmark \\
\hline
\multirow{12}*{AI reasoning} & \cite{sun2023survey} & A survey on a wide range of reasoning tasks and techniques enabled by large foundation models & \cmark & \cmark & \xmark \\
\cline{2-6}
~ & \cite{chen2025harnessing} & A comprehensive survey on the reasoning methods in multiple LLMs collaboration & \cmark & \cmark & \xmark \\
\cline{2-6}
~ & \cite{yu2024natural} & A survey on Natural Language Reasoning (NLR) in text-based tasks, focusing on single-modality unstructured text reasoning & \cmark & \cmark & \xmark \\
\cline{2-6}
~ & \cite{delong2024neurosymbolic} & A survey focusing on neural-symbolic reasoning that combines symbolic logic and neural network learning for knowledge graphs & \xmark & \cmark & \xmark \\
\cline{2-6}
~ & \cite{wang2024exploring} & A first systematic survey on multimodal LLMs reasoning, including evaluation protocols, benchmarks, and classifying reasoning-intensive tasks & \cmark & \cmark & \xmark \\
\cline{2-6}
~ & \cite{li2025perception} & A survey introducing the native multimodal reasoning model (N-LMRM) used for autonomous planning, and the emerging reasoning technologies & \cmark & \cmark & \xmark \\
\cline{2-6}
~ & \cite{patil2025advancing} & A review focusing on the gap between traditional AI reasoning methods and LLM reasoning methods, as well as the challenges posed by the latter & \cmark & \cmark & \xmark \\
\hline
\hline
\end{tabular}
\vspace{-0.4cm}
\end{table*}

\subsection{Related Surveys}

We first position our paper relative to existing surveys and tutorials in the field. Table \ref{tab1} summarizes recent survey papers and perspectives on AI and LLMs for wireless networks or reasoning methods, highlighting their focus and how our work differs. While a few comprehensive surveys have appeared, none to date has specifically focused on reasoning techniques and their layered applications in wireless networking, which is the gap paper aims to fill.

Regarding the surveys on AI for wireless communications and networks, several studies have begun to investigate the intersection of AI/LLMs and wireless networks.  First, many works focus on applying LLMs or AI techniques to network management and operations \cite{zhou2025large}, \cite{long2025survey}, \cite{boateng2025survey}, \cite{zhu2025wireless}. These surveys collectively highlight the potential of LLMs to automate network configuration, monitoring, and optimization tasks. Second, other surveys explore the emergence of large-scale AI models in beyond-5G systems. Chen et al. \cite{chen2024big} discussed the concept of wireless foundation models tailored for 6G networks, and Cui et al. \cite{cui2025overview} presented a broad perspective on integrating AI into 6G architectures, covering system-level visions such as AI-native network design and AI-as-a-service. Also, Golec et al. \cite{golec2025llm} summarized the method of using LLM to detect advanced
persistent threats in 6G networks. Third, some works address techniques for adapting and optimizing AI/LLM usage in communications, such as transfer learning \cite{nguyen2022transfer} and prompt engineering \cite{zhou2025large1}. 

However, all these surveys do not specifically focus on the reasoning methods of LLM in wireless networks, nor do they explore how to utilize these reasoning capabilities across different network layers systematically. Only the \cite{zhou2025large1} mentioned some reasoning methods. This is far from sufficient for the reasoning that is an important key to LLM and has the potential to solve wireless problems. Therefore, we immediately followed up with a summary of the relevant surveys on AI reasoning.

These surveys cover diverse aspects of AI reasoning. The language-centered reasoning and the multimodal LLM reasoning involve two separate surveys, respectively. The former focuses on reasoning in natural language  \cite{sun2023survey}, \cite{yu2024natural}, while the latter pays attention to reasoning capabilities in more widely applied scenarios such as vision \cite{wang2024exploring}, \cite{li2025perception}. Additionally, there are studies on reasoning methods in multi-LLM systems \cite{chen2025harnessing}, which emphasize the collaboration and complementarity among LLMs \cite{luo2025trustworthy}, \cite{luo2025weighted}, \cite{behera2025towards}. Meanwhile, in \cite{delong2024neurosymbolic}, the authors targeted neuro-symbolic reasoning on structured knowledge, bridging logical inference and neural methods for Knowledge Graphs (KGs). Then, Patil et al. \cite{patil2025advancing} studied the trend of reasoning transformation from traditional AI to LLM. Notably, none of these surveys examines reasoning in wireless communications.

To summarize, we have identified the following open issues in the current related surveys:

\begin{itemize} 

\item The current surveys on LLM for wireless networks pay insufficient attention to the reasoning methods. Reasoning is the core of how LLM solves complex network problems, but the existing research mostly focuses on application outcomes or system integration. This makes it difficult to understand the underlying path of LLM in solving network problems, limiting its application expansion in complex scenarios. 

\item These surveys lack specific discussions on each layer of the network. The challenges at different layers of wireless networks (physical, link, network, transport, and application) are significantly different, and the requirements for LLM reasoning capabilities also vary.  Therefore, it is difficult to formulate a hierarchical optimization plan. 

\item The surveys of LLM reasoning lack a deep integration with wireless communication networks. They mostly focus on general fields and fail to consider the particularities of wireless networks, such as dynamic topology and signals. This results in an unclear path for the implementation of reasoning capabilities and a lack of an operational framework.

\end{itemize}

Our survey aims to fill the aforementioned gap and provides the following content: 1) A structured taxonomy of state-of-the-art AI reasoning techniques suited for networking problems; 2) A sequential analysis from each layer of the network and across layers, highlighting how these reasoning technologies address the key challenges of each layer.

\subsection{Our Contributions}
This survey focuses on various layers of wireless communication networks, as well as security issues and cross-layer collaboration, to conduct in-depth discussions on advanced AI reasoning technologies. The advantage is that it can categorize and highlight the challenges and reasoning methods faced by each layer, forming a more practical and applicable panoramic view. After layer-by-layer organization, it is observable how the reasoning chain is executed. This organization can further guide the design of future cross-layer AI architecture and promote the standardization of AI reasoning in wireless networks. The specific contributions are as follows:

\begin{itemize} 

\item We summarize the AI reasoning methods and classify them into three categories: prompt strategies, architectural innovations, and learning paradigms. Moreover, we present the role of reasoning in wireless networks based on the logical framework of ``reasoning method-function-wireless scenario”.

 \item We conduct a detailed analysis of the AI reasoning methods for the physical layer, link layer, network layer, transport layer, and application layer by referring to the network hierarchy. Specifically, we propose the enabling role of AI reasoning at each level by following the framework of ``challenge-AI reasoning solution-typical scenario”. These scenarios represent the cutting-edge development directions of wireless networks.

%\item We also summarize and introduce the AI reasoning methods related to security issues in wireless communication networks and cross-layer collaboration, following the similar approaches in each layer. These two aspects not only affect the overall performance of the wireless network, but also provide a global perspective for optimizing AI reasoning across the entire network.

\item We propose the future research directions for AI reasoning in wireless communications, including agentic AI reasoning, data-efficient learning and generalization, efficient deployment and collaboration of large models, and secure and reliable AI systems. These provide new directions for future research on AI reasoning for wireless communication networks.

\end{itemize}

\begin{figure*}[!t]
\centering
 \includegraphics[width=3.5 in]{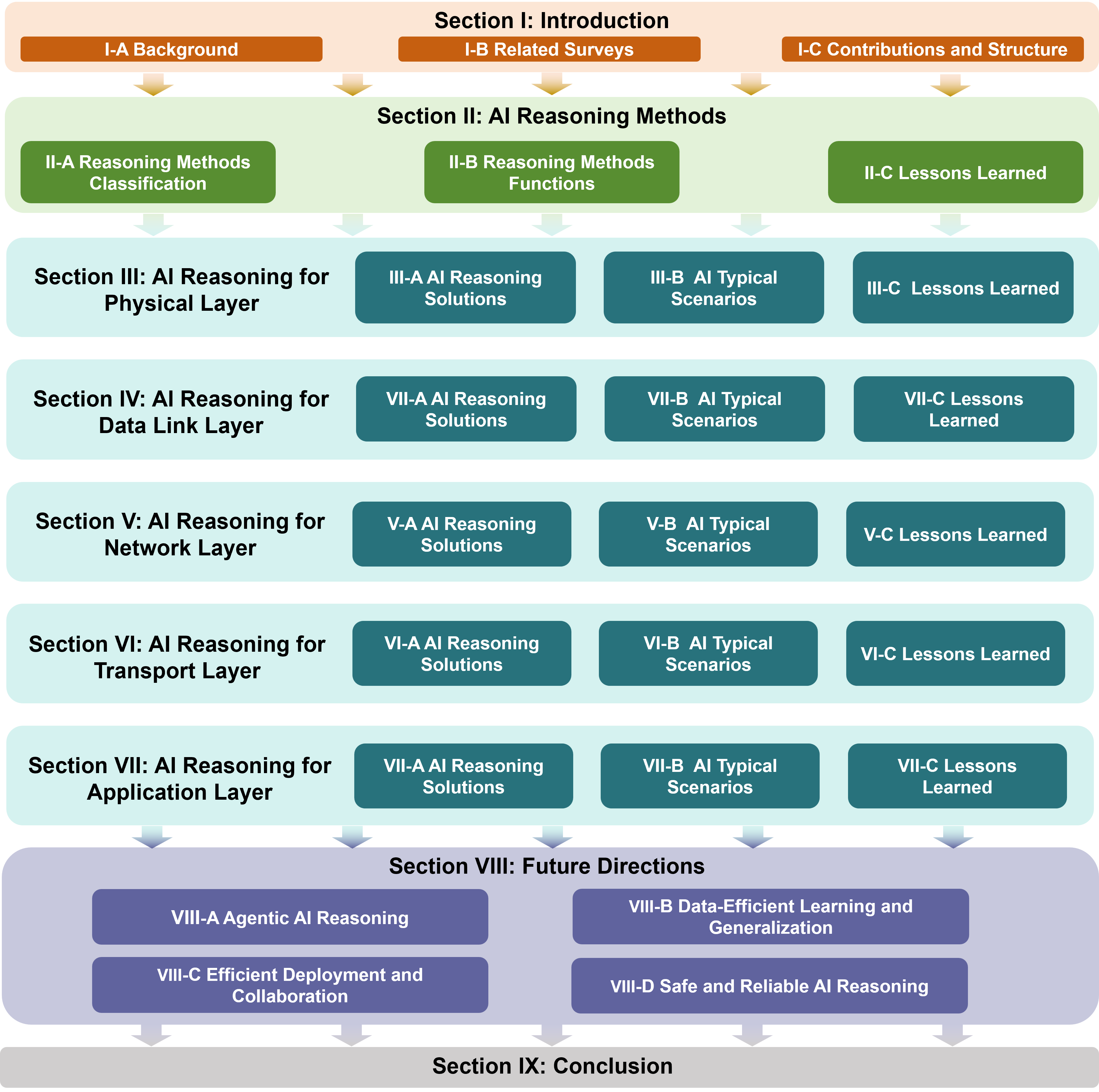}
   \caption{Structure of our survey.}
\label{fig:structure}
\vspace{-0.5cm}
\end{figure*}

Fig. \ref{fig:structure} illustrates the organization of our survey. The remaining content is as follows. Section \ref{sec-II} categorically introduces the AI reasoning methods. Sections \ref{sec-III} to \ref{sec-VII} respectively discuss the enabling role of AI reasoning in the physical, link, network, transport, and application layers. Section \ref{sec-X} further outlines the future research directions. Finally, this survey is summarized in Section \ref{sec-XI}.

\section{AI Reasoning Methods}\label{sec-II}

This section first classifies the AI reasoning methods, then analyzes their functions and applications in wireless networks. In Table \ref{tab:2}, we present the reasoning methods that will be introduced.

\newcommand{\Prompting}{\textcolor{blue}{\scriptsize$\blacksquare$}}    
\newcommand{\Architectural}{\textcolor{green}{\scriptsize$\blacksquare$}}    
\newcommand{\Learning}{\textcolor{red}{\scriptsize$\blacksquare$}}  
\newcommand{\Generation}{\textcolor{red}{\scriptsize\ding{108}}}
\newcommand{\Classification}{\textcolor{purple}{\scriptsize\ding{108}}}   
\newcommand{\Optimization}{\textcolor{teal}{\scriptsize\ding{108}}}     
\newcommand{\Prediction}{\textcolor{brown}{\scriptsize\ding{108}}}    

\begin{table*}[t!]
\centering
\caption{Summary of AI reasoning methods}
\vspace{-0.3cm}
\label{tab:2}
\tiny
\renewcommand{\arraystretch}{1.4}
\begin{tabular}{p{1.9cm}<{\centering}| p{0.4cm}<{\centering}| p{1cm}<{\centering} |p{1.2cm}<{\centering}|p{7.4cm}}
  \hline \hline
\rowcolor{gray!15} % 表头添加浅灰色背景
\textbf{Method} & \textbf{Ref.} & \textbf{Taxonomy} & \textbf{Functionality} & \textbf{Wireless Scenarios} \\
  \hline
CoT & \cite{gao2025langcoop} & \Prompting & \Generation \Prediction & Traffic demand forecasting and sequential resource planning, by step-by-step reasoning)\\ \hline
Self-Consistency  & \cite{wang2022self} & \Prompting & \Generation \Prediction & Ensemble reasoning to improve reliability, e.g., multi-run network fault classification \\ \hline
Tree of Thought               & \cite{yao2023tree} & \Prompting  & \Generation \Optimization& Multi-path planning for complex resource allocation, exploring alternate solution branches \\ \hline
Graph-of-Thought & \cite{yao2023beyond} &\Prompting&\Generation \Optimization& Graph-based topology optimization, modeling network reasoning as graph synthesis\\ \hline
Contextual Reasoning& \cite{liu2025adaptive} & \Prompting &\Generation \Classification \Optimization \Prediction& In terms of resource management, network optimization, and fault diagnosis, etc\\ \hline
Semantic Reasoning& \cite{chiejina2025accord}& \Prompting &\Generation \Classification \Optimization \Prediction& Semantic perception resource allocation and optimization, intelligent interference management, user behavior prediction and traffic optimization, etc\\ \hline
RAG  &  \cite{zhang2024interactive} & \Architectural& \Generation \Classification&  Knowledge-grounded network diagnostics, retrieving configuration manuals or logs for interpretation \\ \hline
ReAct  &  \cite{yao2023react} & \Architectural& \Generation \Classification \Prediction &  Interactive network control with tool/API calls, e.g., querying routing databases during troubleshooting \\ \hline
PAL  &  \cite{gao2023pal} & \Architectural& \Generation&  Automated physical-layer math, e.g., power allocation via code-based solving \\ \hline
Memory-Augmented  & \cite{ko2024memreasoner} & \Architectural& \Generation \Prediction&  Long-term context retention, e.g., traffic pattern prediction\\ \hline
GNN  & \cite{shen2022graph} & \Architectural& \Prediction&  Resource allocation and power control in dynamic environments by modeling the nodes and their connection relationships\\ \hline
Multi-LLM &\cite{fang2025improving} & \Architectural& \Generation \Classification \Optimization \Prediction&  Through model scheduling optimization, avoiding idle or excessive allocation of computing resources on edge servers\\ \hline
Neuro-Symbolic AI & \cite{thomas2023neuro} & \Architectural &\Generation \Classification \Optimization \Prediction & Steady signal detection, especially when dealing with complex semantic interpretation tasks, can maintain logical consistency\\ \hline
Agentic Reasoning & \cite{liu2025lameta} & \Architectural &\Generation \Classification \Optimization \Prediction & Intelligent decision-making, dynamic resource optimization and robustness enhancement in complex scenarios of wireless communication\\ \hline
Self-Refinement& \cite{madaan2023self} & \Learning& \Generation&  Iterative policy refinement, e.g., network configuration tuning\\ \hline
CoT Distillation & \cite{li2024mode} & \Learning& \Generation&  Compact model deployment, such as edge LLMs for network inference\\ \hline
RLHF  & \cite{yue2025does} & \Learning& \Generation \Classification\Prediction&  Aligning LLM outputs to operator intents, fine-tuning on network management feedback\\ \hline
Self-Taught Reasoner& \cite{zelikman2022star} & \Learning& \Generation \Prediction&  Self-improving network optimization, that is, LLM bootstraps its planning via generated reasoning\\ \hline
PickLLM &\cite{sikeridis2025pickllm} & \Learning& \Optimization \Prediction&  Dynamic model selection for network queries, e.g, choosing a lightweight LLM for routine traffic prediction\\ 
\hline
 \multicolumn{5}{l}{Symbols: \Prompting: Prompting; \Architectural: Architectural; \Learning: Learning; \Generation: Generation; \Classification: Classification; \Optimization: Optimization; \Prediction: Prediction}\\
 \hline \hline
\end{tabular}
\vspace{-0.4cm}
\end{table*}

\subsection{Reasoning Method Classification}

\emph{\textbf{1) Prompting Strategy:}} Prompting-based reasoning methods leverage clever prompt designs to elicit step-by-step thinking from a model without modifying its architecture \cite{hong2024advances}. The core idea is to guide an LLM to “think out loud” in natural language, decomposing complex problems into intermediate steps before final answers \cite{li2024think}. For example, CoT prompting provides exemplars or cues that cause the model to generate a sequence of reasoning steps leading to a solution \cite{li2025structured}. It also yields interpretable rationales, as the CoT offers a transparent window into the model’s decision process. Variants like Zero-Shot CoT \cite{kojima2022large}, which simply instructs “Let’s think step by step,” and Few-Shot CoT \cite{kim2023cot}, which provides worked examples. They enable reasoning even without task-specific training for LLM. %This approach can dramatically improve performance on tasks requiring multi-step logic by allocating more computation to harder problems and making the model’s latent reasoning explicit. 

Similarly, Tree of Thought (ToT) prompting expands on this idea by exploring multiple reasoning paths in a search tree \cite{yao2023tree}. At each decision point, the model can branch into alternative thoughts, ultimately converging to the most promising answer. This “deliberative” strategy improves the chance of finding correct or optimal solutions by considering different possibilities (similar to human brainstorming), at the expense of more queries or expanded prompts \cite{boyle2024itot}. Furthermore, there is the Self-Consistency CoT \cite{wang2022self}, which first randomly selects a series of different reasoning paths instead of merely following the most straightforward path in the CoT. Then, it selects the most consistent answer by disregarding the selected reasoning paths.

In general, prompting strategies are useful as they require no model retraining. A single LLM can be prompted to solve many tasks with reasoning. The trade-off is that parsing long reasoning traces increases inference time and token usage, which can be problematic for latency-sensitive applications. Prompting strategies are most appropriate when quick adaptability and interpretability are needed, or when the users want to probe an existing model’s reasoning ability on a new problem without extensive fine-tuning \cite{sun2025survey}. %However, they rely on the model being sufficiently large and capable. Small models might produce flawed or illogical chains, and they may occasionally lead the model astray if the prompt exemplars or instructions are suboptimal \cite{turpin2023language}, \cite{smith2024language}.

\emph{\textbf{2) Architectural Approach:}} In contrast to pure prompting, some reasoning methods introduce architectural or system-level modifications that augment an LLM with external tools, memory, or multi-module workflows \cite{ye2025chatmodel}. The motivation here is to overcome the limitations of standalone LLM reasoning, e.g., mathematical precision, factual recall, or environmental interaction. For instance, ReAct is a paradigm that interweaves reasoning and acting in an interactive loop \cite{yao2023react}. Instead of only producing a verbal CoT, a ReAct agent alternates between thought steps and action commands, such as API calls, tool uses, or environment queries. This allows the model to gather new information and update its plan based on external feedback, as well as to formulate high-level plans that guide which actions to take next\cite{yang2023mm}. The result is a powerful synergy. This model is not limited by its parameter content. It can both explore through actions and explain through thinking. This improves problem-solving breadth and also enhances interpretability and trustworthiness, since one can inspect both the thought trace and the chosen actions \cite{aksitov2023rest}. 

Another architecture-driven method is Program-Aided Language (PAL) modeling \cite{gao2023pal}, which integrates a code execution tool into the reasoning loop. Here, the LLM translates a natural language problem into a Python program, as its CoT, and then offloads the actual computation to a Python interpreter. The logic is that writing code is a more structured form of reasoning for tasks like math or algorithmic questions \cite{yang2025code}. The LLM ensures the problem is broken into solvable steps by generating code, and the interpreter yields a reliable result for each step. This division of labor significantly boosts accuracy on tasks where pure neural reasoning might falter. %For example, PAL with an LLM can solve math word problems more accurately than an LLM using free-form CoT alone, by delegating the execution to a deterministic solver \cite{alex2025pal}. 

Beyond these, architectural innovations include tool-use frameworks, where models call external APIs, search engines, or calculators \cite{shen2024llm}, such as RAG \cite{zhang2024interactive}. Additionally, memory-augmented models can retain long-term working memory \cite{ko2024memreasoner}, and neuro-symbolic systems that combine neural nets with logic rules or KGs \cite{de2025design}. Also, the work involves combining multiple LLMs to collaboratively reason for a specific task \cite{luo2025toward}. Meanwhile, GNNs offer a structured framework specifically designed for processing graph-structured data. Specifically, GNNs can clearly represent entities and their relationships, and support logical reasoning and multi-hop question answering \cite{patil2025advancing}. All share a common design logic, extending the reasoning capabilities of the base model by structural means \cite{haque2025advanced}. The trade-offs often involve added complexity. For example, ReAct may require many interaction steps, and tool-augmented models depend on the reliability and speed of the external tools \cite{lumer2024toolshed}. In scenarios that demand higher precision or access to fresh data, these strategies are more appropriate. They shine when a task can be naturally broken into subtasks like or when no single model can internally capture all the knowledge or skills needed \cite{patil2025advancing}. 

%However, designing these systems requires careful orchestration, deciding when the model should use the tool or how to format tool inputs/outputs. And each added component, including APIs, interpreters, and memory modules can become a point of failure or delay. Nonetheless, for complex decision-making tasks that benefit from grounding in an external reality or formal logic, architectural approaches including ReAct and PAL provide a robust path \cite{karapantelakis2024survey}.

\emph{\textbf{3) Learning Paradigm:}} This method focuses on how the model learns or improves its reasoning abilities \cite{peng2025lmm}. These techniques involve training time or iterative post-processing innovations rather than just prompting techniques. One representative example is Self-Refinement \cite{madaan2023self}, an approach in which the model iteratively improves its output by generating feedback and revising itself. Inspired by how humans edit and refine their answers, a self-refinement loop might have the model produce an initial solution, and critically evaluate it, identifying errors or gaps \cite{xu2024pride}. Then it tries again to produce a better answer, possibly repeating multiple times. LLM itself serves as the generator, the reviewer, and the editor in turn. Such a paradigm blurs the line between training and inference \cite{yang2024idea2img}. The model is essentially learning from its mistakes. Sometimes called test-time learning or meta-reasoning \cite{sui2025meta}. The benefit is a markedly higher answer quality and the ability to correct reasoning flaws dynamically. %Reports show $>20\%$ performance gains on some tasks with GPT-4 through self-refinement vs. one-shot answers \cite{madaan2023self}. 

Another learning-oriented method is CoT Distillation (CoTD) \cite{li2024mode}, which moves the heavy lifting to the training phase. The idea behind CoTD is to teach a smaller or less capable model to perform reasoning by using a larger model’s reasoning traces as training data \cite{zhu2024distilling}. For instance, one can prompt a very large teacher LLM to produce detailed reasoning steps for many problems, then fine-tune a smaller student model on these step-by-step solutions so that it internalizes the reasoning pattern. By distilling the thinking process, the student model can achieve much better reasoning ability than before. This paradigm is powerful for making reasoning widely deployable. Smaller models can be run at the network edge or on devices with limited compute, yet exhibit some of the CoT advantages of the large foundation models \cite{luo2025toward}. Other learning paradigm techniques include fine-tuning on rationales \cite{zhu2025rationales}, where human or machine-generated rationales are added to training sets to instill a habit of explaining.  Reinforcement Learning (RL) for reasoning \cite{yue2025does}, \cite{zhang2025embodied}, it rewards correct and well-justified reasoning paths, akin to how Reinforcement Learning from Human Feedback (RLHF) aligns user preferences. %There is also an ensemble method \cite{fang2024llm}, where the model learns to generate multiple reasoning paths and select the most consistent answer. 

These measures aim to integrate the reasoning ability into the model’s usage pipeline, rather than relying solely on prompting at runtime. However, there exists a trade-off. Training-based methods that require additional data preparation and computing upfront \cite{underwood2023libpressio}. They may sacrifice some flexibility. Once a model is fine-tuned or distilled for a particular style of reasoning, it could be less general than an LLM prompted appropriately \cite{zheng2025review}. However, at inference time, these models can be more efficient, since they might not need extremely lengthy prompts or multi-step interactions.  In summary, training-based reasoning strategies are ideal for situations where reasoning will be inefficiently needed, and we can invest in upfront training to optimize performance. %They trade initial development overhead for swift and reliable reasoning at runtime.

\subsection{Reasoning Method Functions}
 In the wireless networking domain, we can identify four major classes of tasks: generation, classification, optimization, and prediction \cite{khoramnejad2025generative}. 

\emph{\textbf{1) Generation Task:}} These tasks involve producing novel outputs. In wireless networks, this could mean translating a high-level intent into a concrete network configuration or policy, generating semantic content for network management \cite{li2024klonetai}. Here, reasoning techniques can ensure that the generation is logically coherent and constraint-abiding. For instance, CoT prompting has been applied to intent translation problems, where an LLM takes a request like “Provide ultra-reliable low-latency service to Device $X$” and generates a sequence of actions or parameter settings for the network \cite{wang2025chain}. By reasoning step-by-step, the model can interpret the intent, map it to specific requirements, e.g., bandwidth, scheduling priority, coding scheme, and then draft a configuration that meets those requirements \cite{wang2025survey}. This yields an automation workflow that is both transparent and verifiable. Prompting methods are particularly suited for generation because they produce those intermediate “thoughts” that essentially serve as pseudo-code or outlines for the final output.

Additionally, architectural approaches play a role in generation tasks. A tool-augmented method, such as PAL, can generate and execute network configuration scripts. The LLM writes a script to configure the network, and then an interpreter runs it to verify that it indeed produces the desired outcome \cite{mondal2023llms}. This two-step approach catches errors. If the script fails or yields suboptimal results, the reasoning loop can adjust the code. In essence, the combination of LLM and tool acts as an intent fulfillment engine \cite{wang2024netconfeval}, where the LLM’s reasoning ensures all aspects of the request are addressed. Another example is an intent-driven CoT framework that sequentially parses user requests \cite{wang2025chain}, selects reasoning modules, and generates interpretable control policies for a wireless network. Such a framework was shown to successfully take a natural language intent \cite{li2024klonetai}, such as a deployment request, and produce a detailed network configuration plan, with each step justified.

\emph{\textbf{2) Classification Task:}} Classification tasks in networking might include things such as classifying the current network state, identifying interference sources, recognizing user or traffic categories, and others. Traditionally, an ML model might output a class label with limited explanation. Reasoning-augmented AI instead produces a classification along with a rationale, which is useful in wireless domains for validation and trust \cite{zhang2025beyond}. For example, an LLM can be asked not just “Is this link high quality or low quality?” but “Explain whether this link is high or low quality and why.” A prompting strategy can then yield an answer like: “The link is low quality because the Signal-to-Noise Ratio (SNR) is $5$ dB, below the 10 dB threshold, and there have been two handoffs in the last minute, indicating an unstable connection.” This CoT justification allows a network controller to understand the basis of the classification \cite{wang2024know}. Meanwhile, in safety-critical or performance-critical networks, such transparency is crucial \cite{you2025reacritic}. It helps operators verify the decision logic and catch mistakes. For instance, ToT-assisted LLM can identify an interference event and describe the likely cause, such as frequency overlap, weather attenuation, etc., step-by-step \cite{sha2024hierarchical}. This not only yields a classification, but also context. 

Moreover, ReAct can enhance classification-type tasks by allowing the model to query external data before finalizing a decision \cite{karapantelakis2024survey}. For example, consider a fault diagnosis task.  A ReAct agent could perform actions including checking recent logs or reading a knowledge base in between its reasoning, such as whether a router was down, or if a power event occurred \cite{men2025interpretable}. Then, it can decide whether the failure is due to a power interruption or a configuration error. By combining verbal reasoning and tool use, such an approach increases confidence in the classification. Also, architectural methods that maintain a working memory can ensure consistency in decisions \cite{saleh2025usercentrix}. For example, a memory-augmented LLM might remember past classifications to ensure current ones don’t contradict historical trends. Overall, reasoning techniques bring explainability and context awareness to classification.

\emph{\textbf{3) Optimization Task:}}
Wireless networks abound with optimization problems, including allocating channels, scheduling users, etc \cite{liu2025lameta}. These tasks often require strategic reasoning and vision, making them suitable for advanced reasoning methods \cite{thomas2024causal}. Classic optimization algorithms are being complemented or replaced by AI solutions as networks become too complex for simple models \cite{hu2025hybrid}. However, an AI that just outputs an answer without reasoning might miss the chance to explore alternatives or justify its choice. This is where planning-based reasoning techniques shine. ToT, for example, is practically designed for optimization. In a resource allocation scenario, a ToT approach might consider multiple ways to assign resources to users, reason about the interference or throughput in each scenario, prune suboptimal branches, and converge on an allocation that maximizes a utility function. Such a method was reported to drastically reduce computation versus brute force search while achieving near-optimal performance \cite{mousaviinformed}. 

Another example could be an RL augmented CoT where the LLM plans a sequence of network actions and is guided by a reward signal \cite{shokrnezhad2025autonomous}. Effectively, the LLM reasons about how to maximize a reward, which is a form of optimization. The reasoning helps break down the long-horizon planning into feasible steps addressing the complexity incrementally, such as first optimizing coverage, then interference. Empirical results in recent research underscore the benefits: when an LLM was used with CoT reasoning as a planning module for network slicing, it produced interpretable, fine-grained decisions and achieved significant performance gains \cite{boateng2025survey}. These improvements come from the model’s ability to systematically consider constraints and multi-objective trade-offs via reasoning, as opposed to a monolithic end-to-end closed box. Moreover, planning-type reasoning can incorporate domain knowledge or rules at intermediate steps \cite{muppasani2024building}. For instance, during optimization, a reasoning LLM can explicitly recall a rule like “do not allocate adjacent channels to close-by transmitters” as part of its thought process, thereby injecting domain-specific optimality criteria.  In short, reasoning empowers optimization by enabling foresight and structured exploration.

\emph{\textbf{4) Prediction Task:}}
Prediction tasks refer to forecasting future values or events based on current and past data. Common examples in wireless include predicting traffic load in the future, estimating user mobility patterns, or forecasting channel quality for proactive resource allocation \cite{li2024map}. Traditional time-series models or ML regressors can do this, but they often treat the problem as a pure pattern recognition task \cite{mohammadi2024deep}. An LLM with reasoning abilities can approach prediction more like a human expert, by recalling relevant context, accounting for causal factors \cite{felin2024theory}. For instance, models such as ReAct or those with long-term memory architectures can first retrieve network performance data from a certain time when a similar event occurs. Then, they can reason based on the comparison of these conditions with the current situation, thereby predicting the traffic situation for tomorrow \cite{wu2024longmemeval}.

 Moreover, techniques such as CoT can help the model explain its prediction, which is useful for decision-makers \cite{huang2024dtpp}. A predicted value like “Channel quality will drop to $X$” might be accompanied by a CoT: “The user is moving at 50 km/h and rain is expected; high speed plus rain fade typically causes a $5-10$ dB SNR drop, so I predict a drop in quality.” This rationale increases confidence in the prediction. In research, there are examples of using CoT reasoning in multimodal Integrated Sensing and Communication (ISAC) tasks \cite{li2025joint}, \cite{wang2024optimizing}. The model predicts something like a beamforming vector from both radio and visual data in such tasks. In one case, a CoT-enabled multimodal model achieved significantly higher accuracy in beam prediction compared to baseline methods, thanks to its ability to reason through the associations between past radar data, images of the environment, and future channel states \cite{wang2025chain}. % Clearly, if the reasoning chain relies on incorrect assumptions, it could lead to a wrong prediction with a confidently wrong explanation. Thus, ensuring the model’s knowledge is up to date is important \cite{griot2025large}, \cite{che2025eazy}. In summary, prediction tasks in dynamic networks benefit from reasoning methods that inject domain knowledge, long-term context, and interpretability into the forecasting process.

\subsection{Lessons Learned}

The above taxonomy and functional perspectives provide a high-level understanding of how advanced reasoning methods can be leveraged in wireless networking AI, ranging from simple prompt-based CoT to complex tool-using agents. We observe that the choice of reasoning strategy depends on the desired balance of transparency, computational efficiency, and task complexity, as well as the nature of the task (generation, classification, optimization, or prediction) in the network context \cite{feng2025efficient}. Equipped with this understanding, we next delve into protocol-layer-specific reasoning. Examining how these general reasoning techniques map to the unique challenges and decision-making processes at different layers of the network, and how reasoning enables more intelligent and adaptive behavior within each layer’s scope.

\section{AI Reasoning for Physical Layer}\label{sec-III}

The physical layer serves as the foundation of wireless communication architectures, orchestrating signal transmission, modulation/demodulation, and channel coding. The physical layer protocols and mechanisms encounter critical challenges, including channel fading, multi-path interference, and noise. These factors directly impact the performance and reliability of communication systems. 

The physical layer protocols demonstrate several key characteristics \cite{goldsmith2005wireless}. First, these systems exhibit strict real-time requirements that demand signal processing to be completed within milliseconds. Furthermore, these systems operate within dynamic environments where wireless channel conditions vary rapidly across both time and spatial dimensions, making predictability extremely challenging. Additionally, the systems must handle high-dimensional data complexity that involves time, frequency, and spatial domains simultaneously. Moreover, they face stringent optimization constraints that require achieving optimal performance while operating under severe limitations in terms of power consumption, available bandwidth, and hardware capabilities.

\begin{figure*}[!t]
\centering
 \includegraphics[width=5 in]{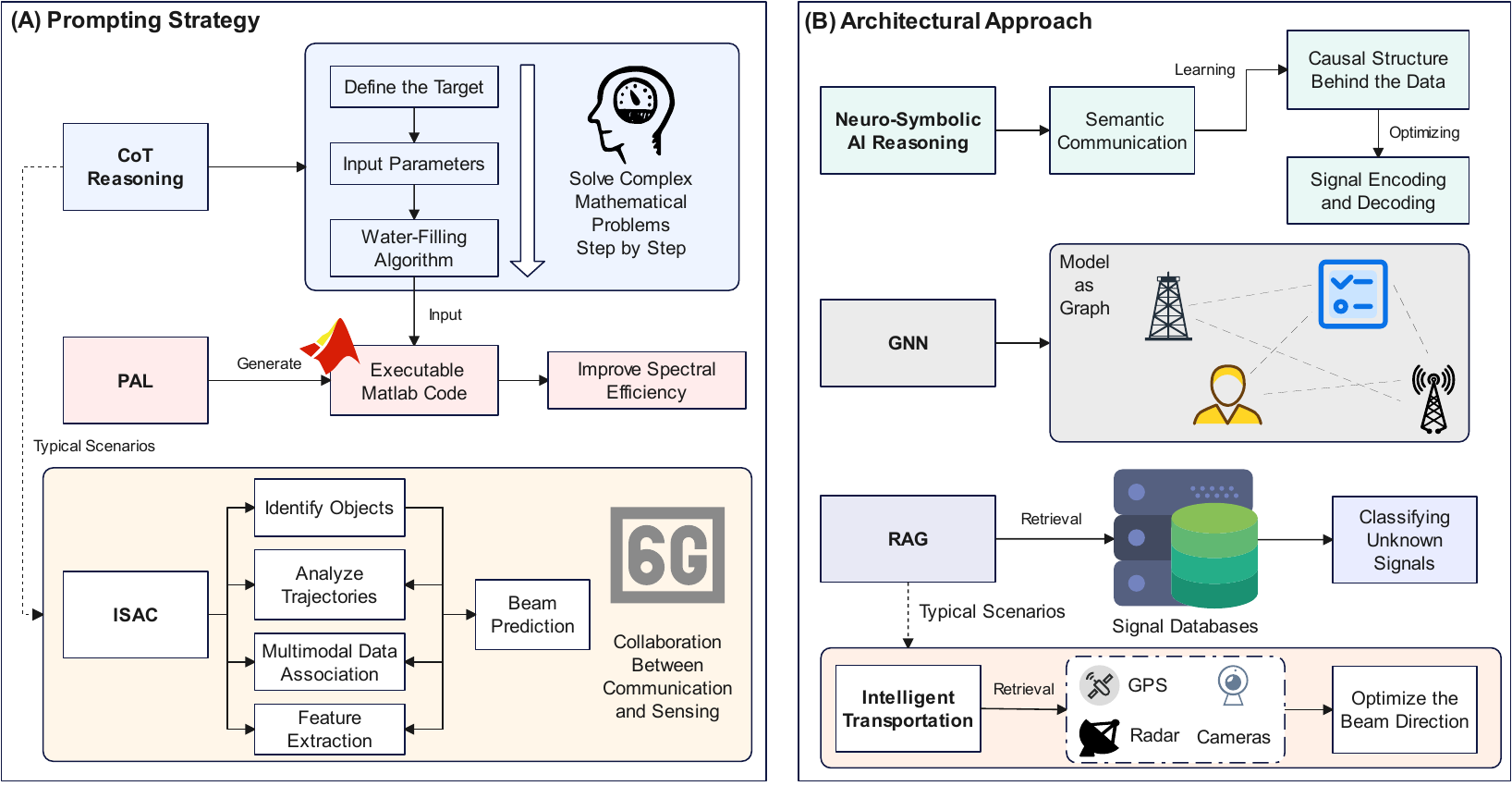}
   \caption{AI reasoning methods in the physical layer. (A) The prompting strategy utilizes CoT reasoning to solve complex mathematical problems, and uses PAL to generate executable MATLAB code to improve spectral efficiency \cite{shao2024wirelessllm}. (B) The architectural approach combines Neuro-Symbolic AI reasoning for semantic communication, optimizing signal encoding and decoding by learning causal structures \cite{thomas2022neuro}. GNN model communication as graphs to enhance understanding and interaction between different elements \cite{shen2022graph}. The RAG method classifies unknown signals by retrieving typical scenarios from extensive signal databases, facilitating intelligent transportation optimization \cite{shao2024wirelessllm}.}
   \vspace{-0.5cm}
\label{fig: AI Reasoning Method in Physical Layer}
\end{figure*}

Consequently, these inherent characteristics render traditional rule-based and mathematical model-based approaches inadequate when dealing with complex real-world scenarios. Thereby, we need the development of more intelligent and adaptive solutions. As illustrated in Fig. \ref{fig: AI Reasoning Method in Physical Layer}, artificial intelligence reasoning methods provide a promising avenue for addressing these challenges in the physical layer implementation.

\subsection{AI Reasoning Solutions}

\textbf{\emph{1) Solutions for Beamforming and MIMO Optimization:}}
Beamforming and MIMO optimization techniques are fundamental for enhancing capacity and coverage in modern wireless communication systems. In massive MIMO and millimeter-wave scenarios, systems must achieve optimal beam design under complex multi-path environments and power constraints. Traditional methods use convex optimization and iterative algorithms \cite{shen2022graph}, such as water-filling and Weighted Minimum Mean Square Error (WMMSE). However, they exhibit significant limitations when dealing with complex non-convex problems, such as high computational complexity, a tendency to converge to local optima, and difficulty in handling discrete constraints.

To overcome these optimization bottlenecks and capture the intricate nature of beamforming problems, recent research has shifted its focus to AI reasoning frameworks capable of decomposing and tackling large-scale tasks. Among them, CoT reasoning and GNN paradigms have become the dominant approaches, delivering near-optimal solutions under strict latency budgets.

\begin{itemize}
    \item \textbf{CoT Reasoning and PAL:} It addresses the high-dimensional optimization challenges in beamforming design through systematic mathematical decomposition \cite{jiang2025comprehensive}. Shao et al. \cite{shao2024wirelessllm} demonstrate the application of a CoT-guided workflow in the OFDM beamforming power-allocation case, splitting the objective of ``maximize sum-rate” into a four-link thought chain: 1) CoT first restates the goal of maximizing aggregate capacity given the total power available across ten tones. 2) It takes as input the per-subcarrier gains and the noise-floor from the channel estimator. 3) It recalls the water-filling algorithm, which allocates power to subcarriers based on their channel quality, prioritizing those with better gains. A bisection loop is then used to find the optimal power distribution that utilizes the total available power. 4) The resulting optimal power distribution is embedded into an LLM-generated, runnable MATLAB cell to generate the exact power pattern for upload to the scheduler. Proof-of-concept runs increase spectral efficiency by $50\%$ at 2.4 GHz. The experiment validates the efficacy of the CoT approach for beamforming and MIMO optimization.

    \item \textbf{GNN:} Wireless communication network optimization is inherently a topological optimization problem for multiple nodes, and GNNs are well-suited to address this class of problems \cite{huang2025graph}. Shen et al. \cite{shen2022graph} proposed a beamforming optimization scheme based on GNN. This approach frames the beamforming problem on a graph to overcome the scalability and generalization limits of prior neural architectures. The wireless systems are modeled as bipartite graphs where nodes represent transmit symbols and antennas, thus reformulating the task as a graph optimization problem. The GNN architecture is designed via deep unrolling, mimicking the message-passing schemes of traditional optimization algorithms. This method yields significant performance gains in Massive MIMO Beamforming, finding near-optimal beamformers for thousands of users in milliseconds and demonstrating robust generalization. A GNN trained on a 50-user network achieves near-optimal performance in a 1000-user network, far surpassing traditional beamforming methods.

\end{itemize}

\textbf{\emph{2) Solutions for Interference Suppression and Signal Detection:}} 
Interference suppression and signal detection are crucial for communication quality in dense wireless networks \cite{adedoyin2020combination}. In these heterogeneous networks, interference is highly complex and dynamic due to limited spectrum and higher network density. Traditional methods like Zero-Forcing (ZF) and Minimum Mean Square Error (MMSE) rely on fixed models and parameters. Owing to their limited adaptability to nonlinear interference and their lack of learning from historical data, these methods exhibit unsatisfactory performance under dynamic conditions.

To overcome the limitations of traditional approaches in interference suppression and signal detection, researchers have recently achieved breakthroughs in various AI reasoning paradigms, mainly in Neuro-Symbolic AI Reasoning.

\begin{itemize}
    \item \textbf{Neuro-Symbolic AI Reasoning:} It combines neural learning capabilities with symbolic reasoning for intent-based semantic communication to realize intent-based semantic communication \cite{thomas2023neuro}. Thomas et al. \cite{thomas2022neuro} pioneered the application of a Neuro-Symbolic AI framework to wireless systems, which leverages it to learn the causal structure behind observed data. The symbolic component is represented by a knowledge base that contains high-level semantic representations, while the neural component, particularly Generative Flow Networks, learns the data's causal structure and the optimal encoding/decoding functions. The learned causal relationships are utilized to optimize signal encoding and decoding. After testing, this method reduced the semantic representation length by $23\%$ and decreased redundant transmissions compared to the implicit semantic communication architecture, while also improving semantic reliability by $63\%$, which significantly suppressed the impact of interference on wireless communication.

\end{itemize}

\textbf{\emph{3) Solutions for Modulation Recognition and Signal Classification:}} Modulation recognition and signal classification are essential for wireless communication systems. These technologies enable spectrum sharing and dynamic spectrum access by identifying modulation formats despite unknown signal parameters. Traditional methods use statistical features and decision theory \cite{al2019performance}, such as maximum likelihood estimation and feature extraction. However, owing to their reliance on extensive prior knowledge and their high sensitivity to SNR and channel variations, they deliver sub-par performance in real-world wireless communication systems \cite{o2016convolutional}.

To more accurately recognize different modulation schemes, an LLM must be equipped with a more comprehensive knowledge base. RAG was specifically designed to address this by augmenting LLMs with domain-specific knowledge to bolster their problem-solving capabilities.

\begin{itemize}
    \item \textbf{RAG Method:} It enhances modulation recognition by combining retrieval mechanisms with generative reasoning capabilities. This approach bears conceptual similarity to traditional look-up table methods \cite{gesbert2007mode}, \cite{loulou2019look}, but demonstrates significantly enhanced intelligence and adaptability. Unlike static look-up tables, RAG dynamically retrieves relevant modulation patterns from large-scale signal databases and leverages generative models to classify unknown signals, thereby addressing the challenge of extensive prior knowledge requirements. In the wireless domain, external knowledge may encompass an extensive array of sources, including device instruction manuals, algorithm textbooks, and wireless standards. The generative component of RAG draws upon this knowledge base to acquire information regarding modulation schemes and channel coding. By leveraging such data, it constructs coherent and contextually relevant content. This, in turn, enhances the LLM’s comprehension of various signal modulation modes. As a result, it enables more precise identification and categorization of unknown channel states. Moreover, based on the identification outcomes, the model can select appropriate demodulation schemes \cite{shao2024wirelessllm}, enhancing the signal demodulation efficiency.
\end{itemize}

\subsection{Typical Scenarios}

\textbf{\emph{1) Integrated Sensing and Communications (ISAC):}}
ISAC is a core vision for 6G communication technology. CoT reasoning can filter out irrelevant semantic information, model spatiotemporal continuity, and enable dynamic cross-modal feature extraction. In an end-to-end ISAC architecture based on a Multimodal LLM (MLLM), the CoT mechanism uses structured instructions and distributed reasoning to decompose complex beam prediction. It progressively correlates the semantics of multimodal data, for example, identifying obstacles in an image and analyzing target interference in radar data. Using this method, the architecture achieved a Top-3 beam prediction accuracy of $70\%$ and $95\%$ in two ISAC tasks, significantly outperforming baseline methods \cite{wang2025chain}.

\textbf{\emph{2) Intelligent Transportation and Vehicular Networking:}}
ENWAR \cite{nazar2024enwar} is a framework that leverages MLLM and RAG method to provide crucial environmental awareness for next-generation wireless communication networks. The framework's core functionality is to integrate multimodal data from various sensors, such as GPS, LiDAR, and cameras. It converts this data into textual descriptions that can be understood by an LLM, which are then used to build a domain-specific knowledge base. For a vehicle's surrounding environment perception, the LLM directly retrieves from this knowledge base and enhances the accuracy of its generated perception results based on the retrieved information. This perceptual capability directly empowers the physical layer of wireless communication. ENWAR can accurately identify obstacles and assess line-of-sight connectivity between vehicles, information that is essential for beamforming and blockage mitigation in high-frequency communications. In terms of performance, the framework demonstrates significant advantages over traditional LLMs, achieving up to $70\%$ relevance, $80\%$ correctness, and $86\%$ faithfulness in its answers. This proves its effectiveness in providing precise interpretations and making decisions in complex, dynamic environments.

\subsection{Lessons Learned}

The application of AI reasoning paradigms, such as CoT reasoning and RAG method, has significantly enhanced the physical layer's performance in core tasks. These methods overcome the limitations of traditional mathematical models, rendering the transmission of raw bit streams in complex and dynamic wireless environments more efficient, adaptive, and reliable. Having thus addressed the key challenges of the physical layer and ensured efficient physical layer data transmission, the subsequent focus is on the key challenges of the data link layer, namely optimizing point-to-point communication within local area networks.

\section{AI Reasoning for Data Link Layer}\label{sec-IV} 

Building upon the reliable point-to-point transmission at the physical layer, the data link layer serves as the bridge connecting to network-level communication protocols. It orchestrates essential functionalities such as medium access control, error detection and correction, and flow control.
In a complex and heterogeneous wireless environment, the data link layer is characterized by numerous challenges, including inter-user interference, dynamic channel conditions, and diverse QoS requirements \cite{plascencia2023comprehensive}. 

\begin{figure*}[!t]
\centering
 \includegraphics[width=5 in]{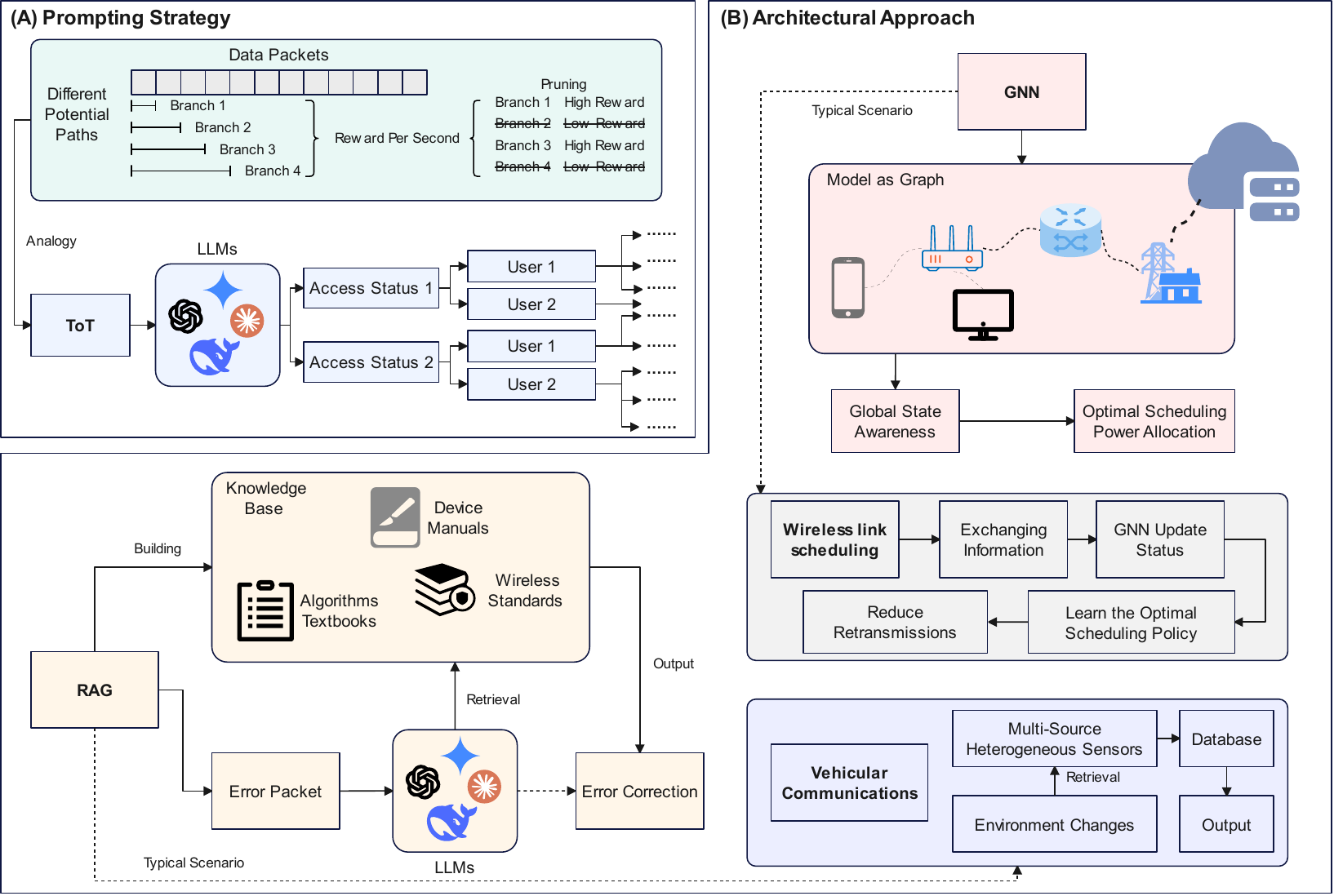}
   \caption{AI reasoning methods for the data link layer. (A) The prompting strategy outlines a method employing a ToT approach to manage data packets for evaluating the access status of multiple users \cite{raviv2019scheduling}. (B) The architectural approach emphasizes the use of GNNs to model communication as a graph, enabling global state awareness and optimal power allocation scheduling \cite{lu2024graph}, \cite{shen2022graph}. Additionally, RAG leverages a knowledge base constructed from device manuals, algorithms, and wireless standards to detail the error correction process in data transmission \cite{robatian2025gec}.}
      \vspace{-0.5cm}
\label{fig: AI Reasoning Method in Data Link Layer}
\end{figure*}

Consequently, these characteristics render traditional fixed protocol designs and static resource allocation schemes inadequate for handling dynamic multi-user scenarios \cite{xie2022resource}, necessitating solutions capable of rapid adaptation to dynamic environments. As illustrated in Fig. \ref{fig: AI Reasoning Method in Data Link Layer}, AI reasoning methods provide an efficient solution to these issues.

\subsection{AI Reasoning Solutions}

\textbf{\emph{1) Solutions for Error Control and Retransmission:}}
The optimization of error control and retransmission algorithms aims to ensure reliable data transmission over unreliable wireless channels by balancing reliability, spectral efficiency, and latency. 
Traditional methods primarily employ fixed coding schemes and static retransmission policies, such as stop-and-wait ARQ \cite{rajhi2020comparison}. They are unable to adapt coding redundancy and retransmission methods to time-varying channel conditions \cite{patil2017dynamic}. And they also lack cross-layer optimization capabilities, making it difficult to align error control with higher-layer application requirements \cite{yigit2019new}. However, current research has ingeniously addressed these issues by leveraging AI reasoning methods like RAG.

\begin{itemize}
   % \item \textbf{GNN:} It has demonstrated remarkable capabilities in addressing issues related to Error Control and Retransmission. Its core concept lies in leveraging the message-passing mechanism within graph structures to optimize error control and retransmission efficiency in communication systems. The JDD-GNN model proposed by Clausius et al. \cite{clausius2025joint} and the MAP-GNN-LDPC model proposed by Nguyen et al. \cite{nguyen2024graph} share the approach of jointly optimizing detection and decoding processes through iterative message passing on factor graphs, thereby reducing error rates and enhancing robustness against interference while eliminating the need for explicit retransmission mechanisms. Additionally, Lee et al. \cite{lee2021decentralized} employed GNNs to optimize selective retransmission strategies in distributed inference by triggering retransmission only for nodes with high misdetection or high outage probabilities. Combined with maximum ratio combining (MRC) techniques, this significantly reduces communication overhead and error rates. Overall, GNN fundamentally enhances the fault tolerance and retransmission efficiency of communication systems through efficient message-passing and inference mechanisms.
    
    \item \textbf{RAG Method:} An approach for Automatic Speech Recognition (ASR) error correction, proposed by Robatian et al., \cite{robatian2025gec} utilizes RAG. The core of this methodology involves first constructing a knowledge base that pairs ASR predictions with their corresponding ground truths. When a new ASR transcription is generated, the system retrieves the most similar examples from this knowledge base using metrics such as Term Frequency-Inverse Document Frequency (TF-IDF) vectors and cosine similarity. These retrieved examples are then used as input for In-Context Learning (ICL), which enables an LLM to generate a corrected transcription based on identified error patterns. This paradigm can be analogously applied to error control and retransmission in wireless communication. A channel coding knowledge base can be established, storing typical data packet error patterns that occur under various channel conditions, along with their corresponding correct packets. Upon receiving a corrupted packet, the receiver would not immediately request a retransmission. Instead, it would first employ a similarity metric to retrieve the ``error pattern” from the knowledge base that is most similar to the corrupted packet. Subsequently, a lightweight generative model would utilize this ``error pattern” and the retrieved ``correct data” as input to intelligently infer and generate the original, correct data packet. This ``retrieval-augmented correction” paradigm has the potential to reduce the frequency of full retransmissions, thereby enhancing data transmission efficiency and system throughput.
\end{itemize}

\textbf{\emph{2) Solutions for Link Adaptation and Rate Control:}}
It requires the dynamic adjustment of transmission parameters based on instantaneous channel conditions to maximize spectral efficiency and ensure reliability.
Traditional methods primarily rely on threshold-based switching mechanisms \cite{saha2018link} and static lookup tables \cite{ramezani2021cqi}. These methods are static, and they struggle to cope with non-stationary channel statistics and interference patterns, and are also unable to adjust strategies in real-time to adapt to dynamic environments. Therefore, current researchers are beginning to explore more dynamic AI reasoning methods to solve these problems.

\begin{itemize}
    \item \textbf{GNN:} GNNs first model wireless networks as graphs. In this graph, nodes can represent users, antennas, or access points, while edges denote their connections or interference relationships. These node and edge features are used as inputs to the GNN. Lu et al. \cite{lu2024graph} propose that by maximizing key metrics such as the total network throughput and rate, GNNs can predict throughput performance under different network configurations and provide a basis for link adaptation decisions. Regarding rate control, Shen et al. \cite{shen2022graph} demonstrate how GNNs employ distributed message-passing mechanisms by exchanging messages between nodes to learn and output optimal power allocation schemes.   This approach makes GNNs an ideal choice for rate control in dynamic wireless environments.   In summary, the powerful learning and generalization capabilities of GNNs provide both theoretical foundations and practical frameworks for addressing core issues in wireless networks.
    
    \item \textbf{RAG Method:} RAG addresses the challenges of link adaptation and rate control by leveraging external knowledge sources to bridge the gap between domain-specific context and general knowledge.  In wireless scenarios, RAG retrieves information from external databases, such as device manuals, algorithm textbooks, and wireless standards, to construct coherent and contextually relevant content. This enables RAG to enhance the understanding of relationships among physical components in wireless systems. For example, RAG can utilize knowledge about modulation schemes and channel coding to perform more accurate link adaptation in response to varying channel states \cite{shao2024wirelessllm}. Furthermore, with its adaptive feedback mechanism, RAG continuously refines its outputs based on new data, user interactions, and updated protocols, ensuring high accuracy and credibility for knowledge-intensive tasks like link adaptation and rate control.
\end{itemize}

\textbf{\emph{3) Solutions for Multi-User Access and Scheduling:}}
Multi-user access and scheduling aims to efficiently coordinate simultaneous multi-user access to maximize system throughput and fairness. This is a particularly challenging problem in modern wireless networks with a proliferation of devices and diverse application requirements. 
Traditional methods such as Time Division Multiple Access (TDMA), Frequency Division Multiple Access (FDMA), and Code Division Multiple Access (CDMA) \cite{castaneda2016overview}, \cite{liu2024road} rely on pre-determined resource allocation schemes. This fixed mode is unable to adapt to dynamic changes in user demands and channel conditions, leading to low resource utilization and a lack of intelligence in handling the dynamic QoS requirements of different users \cite{xie2022resource}. Therefore, emerging AI reasoning methods, such as ToT, offer promising solutions to more effectively address this issue.

\begin{itemize}
    \item \textbf{ToT Reasoning:} While no research directly applies the ToT framework to multi-user access and scheduling problems, some existing methods share a similar core design philosophy with it. The method proposed by Raviv et al. \cite{raviv2019scheduling} exemplifies this approach, relying on a decision-making process that involves multi-path exploration, proactive evaluation, and pruning.  Unlike traditional single-path algorithms, Reward Per Second (RPS) is more exploratory. It explores a limited ``window” of $K$ packets, generating multiple potential decision paths. For each packet, RPS proactively evaluates its ``Reward Per Unit Time” to quantify each path's value. It then prunes unviable paths by assigning a zero reward to packets that will miss their deadlines.  Finally, RPS selects the packet with the highest ``reward per unit time” from the window for transmission.  This multi-step, quantitative evaluation process is fundamentally similar to the ToT framework, which advances a decision by evaluating multiple branches and selecting the most promising path.  In the future, abstracting complex decision logic from algorithms like RPS into ``thoughts” and leveraging the inference capabilities of LLMs could lead to more efficient and flexible optimization for these problems.
   % \item \textbf{GNN:} He et al. \cite{he2021overview} proposed that the core idea of using GNNs to optimize multi-user access scheduling is to model the communication links in wireless networks and their mutual interference relationships as graphs. It then learns from the structural information of this graph to make efficient scheduling decisions. Specifically, one method involves first constructing a Device-to-Device (D2D) network as a graph and using graph embedding techniques to compute low-dimensional feature vectors for each node, which represent its communication quality and interference environment. A classifier can then learn the scheduling policy based on these features, an approach that does not require precise channel information.  Furthermore, in ultra-dense millimeter-wave networks with highly complex interference, researchers also utilize spatial-based Graph Convolutional Neural Networks (GCNs) combined with primal-dual learning to achieve efficient scheduling of multi-user access by having the GNN learn and decide which links should be temporarily deactivated.% This GNN-based approach outperforms traditional scheduling strategies in terms of both speed and accuracy.
\end{itemize}

\subsection{Typical Scenarios}

\textbf{\emph{1) Wireless Link Scheduling:}} Lee et al. \cite{lee2021decentralized} proposed modeling a wireless network topology as a graph. The objective of the link scheduling problem is to determine which edges in the graph can be activated simultaneously to maximize network throughput while avoiding interference. Each device in the graph needs to decide its state and link quality based on its own features and those of its neighbors. In the inference phase, each device uses its local GNN model and exchanges information with its neighbors to make decentralized decisions about its link activation state. In this manner, GNNs enable devices to make real-time, decentralized link scheduling decisions without a central controller, thereby avoiding issues such as severe traffic congestion, computational limitations, and single points of failure that can occur in large-scale wireless systems.% Experiments demonstrated that the GNN method reduced the average number of retransmission rounds required by $42\%$ to $52\%$.

\textbf{\emph{2) Vehicle-to-Vehicle (V2V) communication:}} Mohsin et al. \cite{mohsin2025retrieval} proposed using a RAG framework to integrate multimodal data, thereby optimizing applications for the multi-user access in V2V communication. The framework leverages real-time data from diverse sensors to provide a unified and comprehensive environmental awareness for an LLM. First, camera images are converted into detailed text descriptions using an image-to-text model (such as GPT-4o). Then, GPS data is used to calculate the distance and azimuth between vehicles. Finally, LiDAR and YOLO object detection models are employed to identify and count vehicles and other obstacles within the images. This information is also added to the text description. All of this pre-processed textual information is stored in a vector database. When a response for wireless channel access control is required, the system performs a vector search to retrieve the most relevant context from the knowledge base. This retrieved information, combined with a structured prompt, is used to generate the final response, enabling the LLM to make real-time, environment-based decisions.

\subsection{Lessons Learned}

AI reasoning paradigms such as GNN, ToT, and RAG have shown strong performance in the Data Link Layer. They enable rapid and efficient scheduling for large-scale user access and ensure a low bit error rate and effective retransmission during communication.  Building on this, we will now focus on the Network Layer and discuss how AI reasoning methods can be applied.

\section{AI Reasoning for Network Layer}\label{sec-V}

While the data link layer addresses the efficiency of local, single-hop communication, the network layer serves as the pivotal orchestrator on a global scale. It is responsible for end-to-end data transmission path planning, global resource coordination, and network topology management \cite{zhang2022topology}. 
Within the increasingly intricate wireless networks, the network layer confronts unprecedented challenges, including encompassing dynamic routing selection, topology reconfiguration, and network virtualization across multiple dimensions. These challenges, characterized by their highly dynamic nature, high-dimensional complexity, and strict real-time requirements, render traditional simplified rule-based models less effective in such scenarios \cite{tang2025multi}.

\begin{figure*}[!t]
\centering
 \includegraphics[width=5 in]{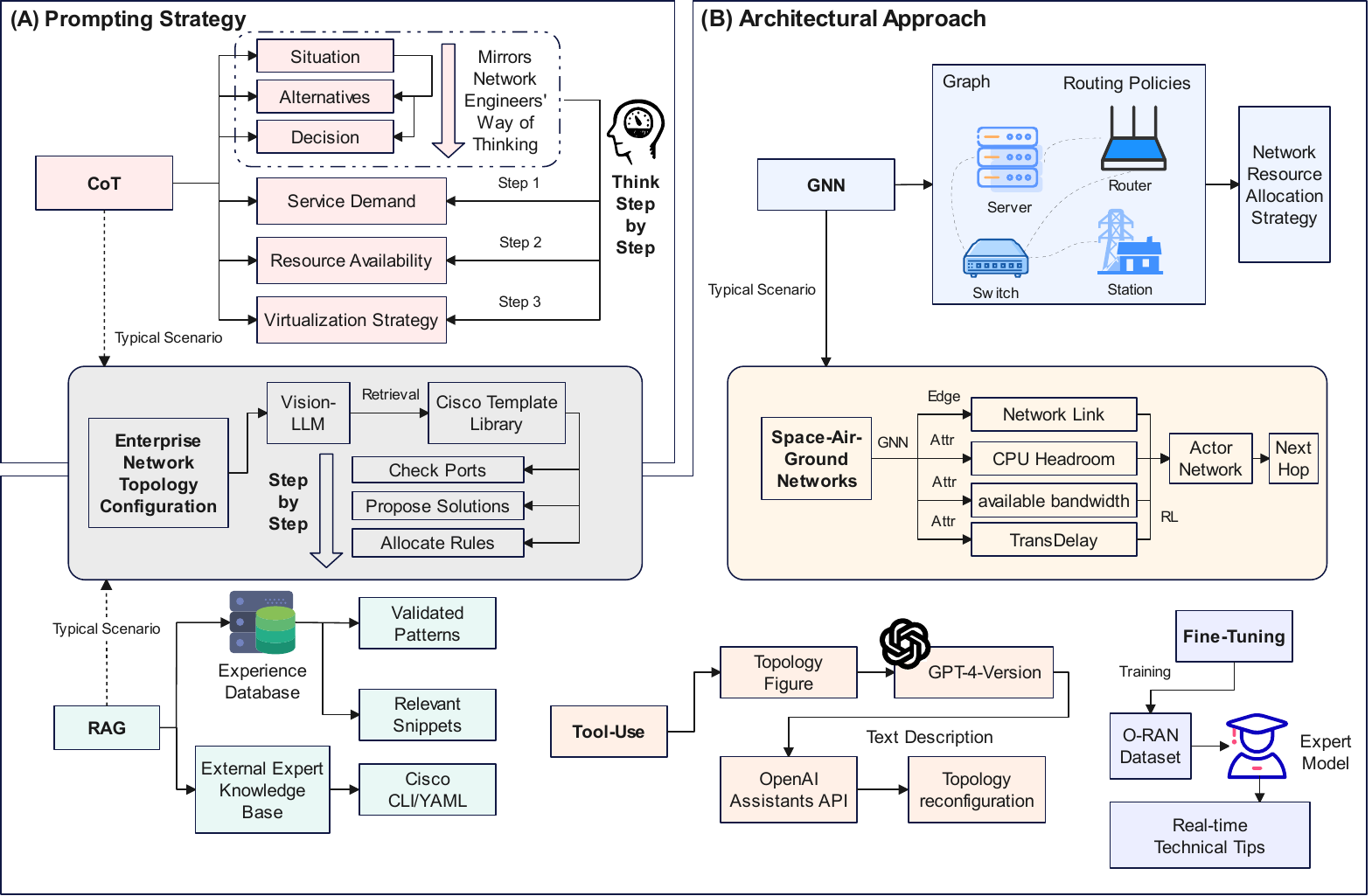}
   \caption{AI reasoning methods for the network layer. (A) The prompt strategy employs CoT reasoning to reflect the cognitive process of network engineers through a systematic three-step methodology: situation analysis, evaluation of alternatives, service demand assessment and decision-making, resource availability evaluation, and virtualization strategy development \cite{dandoush2024large}, \cite{djuhera2024scott}. (B) The architectural approach leverages GNNs to process network topology information, enabling intelligent routing strategies and network resource allocation policies \cite{zhang2024optimizing}. Furthermore, the tool-use method employs GPT-4-Vision to generate textual descriptions from topology images, enhancing the multimodal capabilities for topology reconfiguration \cite{ifland2024genet}. Fine-tuning is conducted using the O-RAN dataset to refine the model into an expert-level model, providing more specialized recommendations \cite{lotfi2025oran}. Additionally, RAG incorporates the experience database to offer empirical guidance for topology design and network management \cite{zhao2023confpilot}.}
         \vspace{-0.5cm}
\label{fig: AI Reasoning Method in Network Layer}
\end{figure*}

To address these challenges, researchers have recently explored a wide range of AI reasoning methods, as shown in Fig. \ref{fig: AI Reasoning Method in Network Layer} hoping to adapt to the emerging demands of the network layer.

\subsection{AI Reasoning Solutions}

\textbf{\emph{1) Solutions for Network Topology Reconfiguration and Node Scheduling:}}
Network topology reconfiguration and node scheduling are vital network layer challenges. They involve dynamically adjusting network structures and optimizing node operations to match traffic demands and node conditions. With the rise of green and intelligent networks, the focus has shifted to maximizing energy efficiency while maintaining service quality \cite{banerjee2020self}. 

Traditional topology management relies on a ``peak capacity planning” model, which uses static configurations to handle maximum load. While this guarantees performance during peak times, it leads to significant resource waste during off-peak periods, and fails to handle the inherent spatial and temporal non-uniformity of network traffic \cite{banerjee2020self}. As a result, researchers are now actively exploring the integration of AI reasoning methods, such as tool-use and RAG, to address these challenges and achieve intelligent, automated network operations.

\begin{itemize} 
    \item \textbf{Tool-Use Approach:} GeNet \cite{ifland2024genet} used a tool-use approach to solve network-layer topology reconfiguration problems. This framework achieves its functionality through the collaboration of two core modules. The topology understanding module utilizes GPT-4 as a tool to convert an input network topology image into a detailed textual description. The intent realization module employs the OpenAI assistants API as a tool to execute more complex tasks. It receives the user's natural language intent and performs reasoning based on the topology's textual description and the device configuration files. The task of this tool is precisely defined: if a topology change is required, it updates the topology's textual description file; simultaneously, it configures all newly added components and updates the device configuration files accordingly. By working in series, these two modules enable GeNet to first understand the network topology from an image and then update the topology's textual description and device configurations based on the user's intent. %In experiments, GeNet successfully configured 79\% of the intended components and 72\% of the newly added components, significantly reducing the workload for engineers and accelerating the network design process.
    \item \textbf{RAG Method:} Zhao et al. \cite{zhao2023confpilot} proposed using an RAG approach to optimize network topology reconfiguration. The core idea of their method is to combine a traditional end-to-end generative model with an external knowledge base to enhance the accuracy and scalability of command generation in resource-constrained environments. First, CONFPILOT utilizes a parser to extract semi-structured configuration information from various vendor device manuals. This information is then standardized into a JSON format and stored in a searchable Configuration Library, which provides the model with a rich source of external knowledge. Next, in the syntax retrieval stage, upon receiving a natural language configuration intent, CONFPILOT's retriever identifies and precisely matches the most relevant command syntax from the ConfLib. Finally, in the command generation stage, the generator takes the user's intent and the retrieved syntax as input and uses a pointer-generator network to produce the final command sequence. By learning to generate configuration commands directly from device manuals, this framework significantly reduces the manual effort of engineers in consulting documentation, which greatly enhances configuration efficiency.
\end{itemize}

\textbf{\emph{2) Solutions for Network Slicing and Virtualization:}}
Network slicing and virtualization are crucial for creating isolated virtual network instances on shared physical infrastructure, delivering customized services to different users and applications. The growth of diverse industries has made it essential to achieve both efficient resource sharing and strict performance isolation \cite{abdellatif2022dynamic}. 

Traditional virtualization, often tied to specific hardware and static resource allocation, is limited in its ability to handle varied service demands. These methods lack the flexibility for rapid network function deployment and fail to provide guaranteed performance isolation, which is particularly challenging in multi-tenant environments where true resource collaboration is difficult \cite{bikkasani2024ai}. To address the issues mentioned above, researchers have proposed various AI reasoning strategies, such as RAG, and fine-tuning.

\begin{itemize} 
  %  \item \textbf{CoT Reasoning:} Network slicing and virtualization is a complex problem that requires comprehensive consideration of multiple issues, such as resource allocation, security, and QoS. The ``multi-step decomposition, gradual solution” approach of the CoT method is highly compatible with these characteristics. Dandoush et al. \cite{dandoush2024large} proposed a solution for network slicing and virtualization using CoT: 1) Through a prompt engineering strategy, an LLM is guided to translate the user's natural language intent into specific technical requirements. For example, the intent ``high-quality video calls for remote medical care” is translated into slicing requirements for high throughput and low latency. 2) The LLM is guided to map these technical requirements to the available network infrastructure and generate standard deployment description files. 3) An LLM-assisted network agent coordinates the deployment, monitoring, and dynamic resource adjustment of the network slices based on these description files. This method can significantly accelerate the entire network service configuration process, substantially improve the end-user experience, and enable network operators to quickly adapt to changing market demands.
    \item \textbf{RAG Method and Fine-Tuning:} Lotfi et al. \cite{lotfi2025oran} proposed the ORAN-GUIDE framework, which uses RAG and fine-tuning to address network slicing challenges in Open Radio Access Networks (O-RAN). The framework's core is its dual-LLM architecture, which decouples domain knowledge generation from task execution. First, a lightweight LLM called ORANSight is trained on O-RAN-specific data, such as slicing policies and configuration logs. This model acts as a ``domain expert”, receiving structured network state snapshots in real-time and dynamically generating technical prompts. This process follows a RAG paradigm, aiming to combine learned system knowledge with real-time observational data. Next, the prompts generated by ORANSight are fused with learnable prompts and input into a fixed, general-purpose GPT model. This GPT model translates these prompts into high-level semantic representations, which are used to augment a DRL agent. The DRL agent then leverages these representations to guide its decisions on network slicing. This framework overcomes the low sample efficiency and weak generalization challenges faced by traditional DRL methods in dynamic, high-dimensional O-RAN environments, providing a unified approach for intelligent decision-making in complex wireless systems.
\end{itemize}

\textbf{\emph{3) Solutions for Dynamic Routing Selection:}} 
Routing is a core challenge in the network layer, focused on selecting optimal data transmission paths \cite{dai2021routing}. The rise of mobile devices and dynamic topologies in modern wireless networks creates a need for smarter routing. 

Traditional protocols such as OSPF and BGP, which use static link weights or hop counts, fail to model multi-dimensional link quality and struggle with multi-objective optimization \cite{dai2021routing}. However, routing selection is a multi-step problem that can be effectively addressed by CoT reasoning, making it highly suitable for optimization using CoT.

\begin{itemize} 
    \item \textbf{CoT Reasoning:} Djuhera et al. \cite{djuhera2024scott} proposed a solution for dynamic routing planning in wireless communication. Their approach, called Strategic CoT (SCoT), utilizes CoT to decompose complex problems into a series of intermediate reasoning steps: 1) Coarse Path Generation. The LLM quickly generates a rough, feasible initial routing based on high-level user commands and basic environmental information, intentionally overlooking some fine-grained wireless communication constraints. 2) Region-of-Interest Refinement. In this phase, the LLM focuses on critical areas within the generated path. By analyzing detailed wireless environment data in these regions, it optimizes the routing to avoid areas with poor communication quality while ensuring major performance requirements are met. 3) Fine-Grained Optimization. This final stage involves the LLM making more granular adjustments. It considers all remaining complex constraints, such as real-time traffic, energy consumption, and latency requirements, to fine-tune the routing path for optimal communication performance and resource utilization efficiency. This phased, structured reasoning approach allows SCoT to reduce the search space for dynamic planning.%, leading to a computational time reduction of up to 62\% in wireless-aware path planning experiments while maintaining near-optimal path gain.
\end{itemize}

%\textbf{\emph{4) Solutions for Optimization of Global Network Resources:}}
Optimizing global network resources is a fundamental challenge in the network layer. The goal is to efficiently allocate a wide array of resources, including bandwidth, computational power, storage, and energy, to maximize overall performance \cite{abdellatif2022dynamic}. With the proliferation of 5 G/6 G, these resource optimization problems have become far more complex.

\subsection{Typical Scenarios}

\textbf{\emph{1) Dynamic Service Function Chain (SFC) Mapping in Space-Air-Ground Networks:}}
In large-scale SAGIN environments characterized by dynamic topology and heterogeneous resource constraints, Graph Pointer Neural Network (GPNN) combined with policy-gradient reinforcement learning have been leveraged to optimize SFC deployment \cite{zhang2024optimizing}. The GPNN encoder extracts spatio-temporal features from the network graph, including node CPU, link bandwidth, and propagation delay, while the RL agent learns a stochastic policy that jointly minimizes end-to-end latency and maximizes resource utilization. Notably, the GPNN’s graph-convolution layers enable global awareness of inter-satellite and terrestrial links, allowing the agent to adaptively reroute SFCs when LEO satellites drift or gateway congestion occurs. %Through iterative interaction with the SAGIN simulator, the system discovers mapping strategies that outperform three baselines (MLRL, NFVdeep, RL) by $10.17\%$ in long-term reward/cost ratio, $15.34\%$ in node utilization, and $16.38\%$ in request acceptance rate. 

\textbf{\emph{2) Enterprise Network Topology \& Configuration Co-Pilot:}}
In enterprise campus networks where engineers must rapidly respond to intents such as ``add three PCs to Lab-3 with full Internet reachability”, a multimodal LLM co-pilot named GeNet integrates CoT and RAG to jointly update topology diagrams and device configurations \cite{ifland2024genet}. Given a topology image and a natural-language intent, GPT-4 Vision first converts the diagram into a structured textual graph, then the system retrieves candidate configuration snippets from a curated Cisco template store. Via explicit CoT reasoning ``identify switch port exhaustion, propose daisy chain switch, assign new subnet”, GeNet generates topology edits and IOS commands that achieve $92\%$ successful re-configuration and $72\%$ correct setup of newly added components at low temperature.%, while maintaining deterministic outputs at low temperature. %This approach reduces mean design time from 45 minutes to under 3 minutes, demonstrating the practicality of neuro-symbolic co-pilots in production NetOps workflows.

\subsection{Lessons Learned}

AI reasoning has elevated the network layer's ability to tackle dynamic routing, topology reconfiguration, slicing, and global resource allocation with greater adaptability, accuracy, and efficiency than conventional heuristics. Those techniques can cascade to fine-grained transport-layer mechanisms, setting the stage to jointly optimize congestion control and end-to-end reliability across the entire protocol stack.

\section{AI Reasoning for Transport Layer}\label{sec-VI}

As a key component in wireless communications, the transport layer ensures end-to-end data scheduling, congestion control, and reliable delivery. Positioned between the network and application layers, it directly impacts throughput, stability, and user experience (QoS/QoE). With rising user density and diverse service demands, especially in latency-sensitive scenarios such as Multi-access Edge Computing (MEC), AR/VR, and industrial IoT, the transport layer must intelligently coordinate upper-layer requirements with fluctuating lower-layer link conditions.

\begin{figure*}[!t]
\centering
 \includegraphics[width=5.5 in]{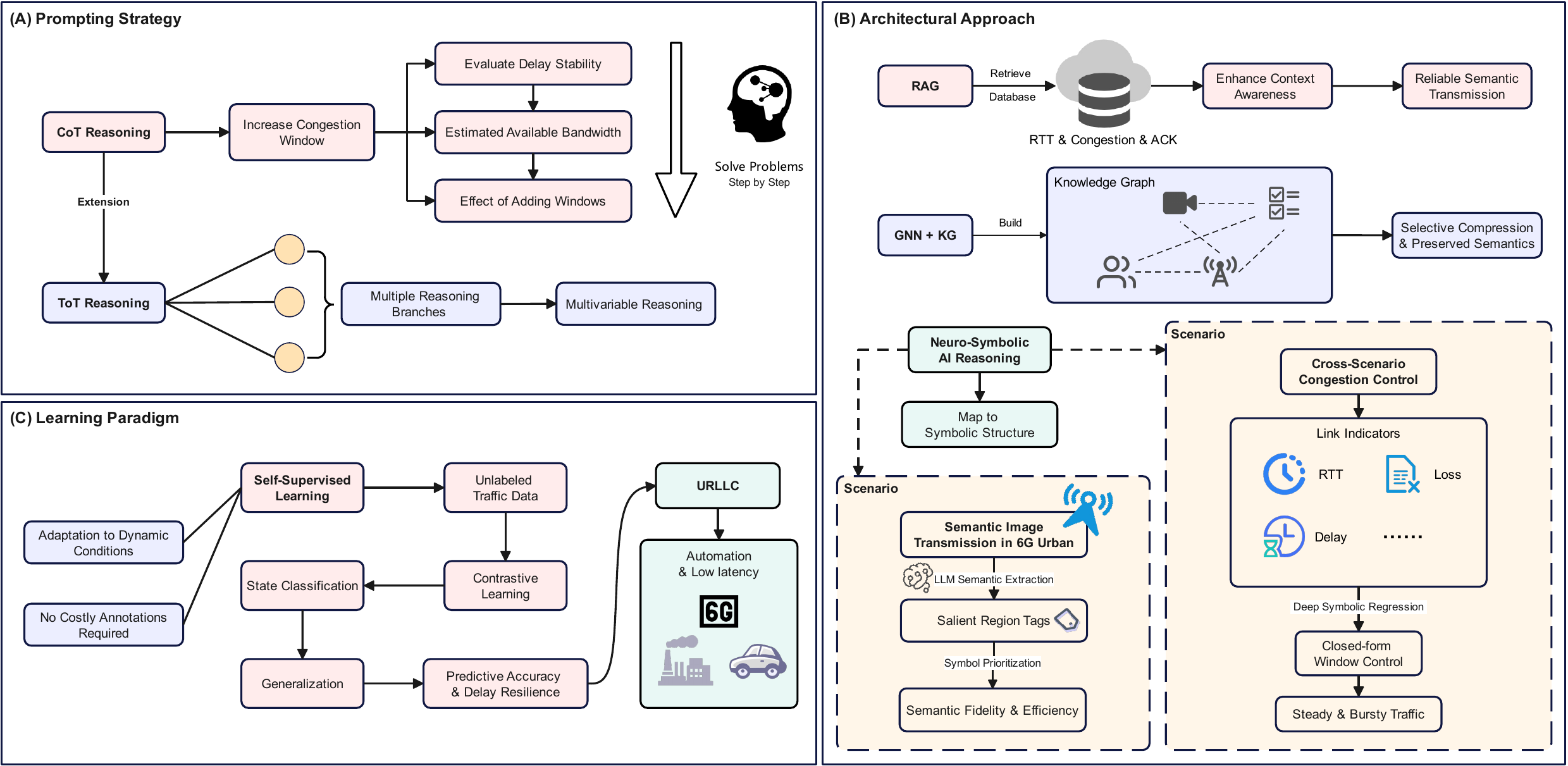}
   \caption{AI reasoning methods for the transport layer. (A) Prompting strategies such as CoT and ToT support step-by-step and multibranch reasoning \cite{yao2023tree} \cite{wei2022chain}, enabling adaptive congestion control. (B) Architecturally, RAG retrieves historical signals for context \cite{lewis2020retrieval}, neuro-symbolic methods enable symbolic logic inference \cite{martins2024closed}, and GNN+KG improves semantic structuring \cite{salehi2025llm}. (C) In learning paradigms, SSL helps models adapt to unknown network states \cite{achiam2023gpt}.}
     \vspace{-0.5cm}
\label{fig: AI Reasoning Method in Transport Layer}
\end{figure*}

The wireless transport layer exhibits distinct characteristics that make it well-suited for AI reasoning-based optimization. First, it involves strong temporal dependencies and causal chains, with decisions heavily influenced by sequential feedback such as Round-Trip Time (RTT), Acknowledge character (ACK) timing, and packet loss \cite{wei2022chain}. Second, the layer operates under high dynamics, with link quality and bandwidth fluctuating frequently, which limits the effectiveness of static strategies. Lastly, decision-making in this context requires the fusion of multi-source data across layers, and the ability to generalize across various topologies and deployment contexts \cite{bennis2018ultrareliable}. These features define the transport layer as a complex sequential decision task, making it highly compatible with AI reasoning techniques, including causal modeling, adaptive policy learning, and real-time inference. Figure \ref{fig: AI Reasoning Method in Transport Layer} illustrates AI reasoning methods applied to the transport layer.

\subsection{AI Reasoning Solutions}
\textbf{\emph{1) Solutions for Congestion Control:}} Congestion control has become a critical issue in wireless environments with the advent of 5G and the shift toward 6G, driven by increasing user density, traffic volume, and pressure on limited bandwidth resources \cite{she2020deep}. Dynamic traffic patterns and unpredictable channel conditions necessitate mechanisms that balance link utilization and packet loss to ensure energy efficiency and reliable delivery \cite{alipio2025cache}. Traditional algorithms, such as standard TCP, struggle in these environments, as they assume stable conditions, fail to estimate available bandwidth effectively, and react slowly to changes \cite{jiang2025combining}. This often results in high latency, packet loss, reduced throughput, and inefficient resource utilization, as demonstrated by studies comparing TCP performance in wireless channels \cite{jiang2025combining}, \cite{tang2022tcp}, \cite{balakrishnan2002comparison}.

To overcome these limitations, researchers have directed efforts toward AI reasoning paradigms tailored to transport-layer adaptation. This emerging direction encompasses methods such as CoT and ToT reasoning, along with neuro-symbolic reasoning, all of which introduce structured mechanisms for more effective congestion control.

\begin{itemize}
    \item \textbf{CoT and ToT reasoning:} CoT and ToT reasoning support interpretable decision-making in dynamic transport environments. CoT enables LLMs to analyze time-series metrics such as RTT and packet loss in a sequential manner, breaking down complex actions, such as congestion window adjustment, into logical steps \cite{wei2022chain}. ToT further enhances this capability by evaluating multiple control strategies in parallel, including aggressive pacing, conservative pacing, and adaptive variations, and then selecting the most appropriate one based on simulated utility scores \cite{yao2023tree}. These structured reasoning processes are consistent with transport protocol design and significantly improve responsiveness to wireless conditions.
    \item \textbf{Neuro-Symbolic AI Reasoning:} It integrates symbolic logic and neural representations to model packet behaviors, such as retransmissions, queue buildup, and ACK delays \cite{martins2024closed}. Specifically, this method begins by training a reinforcement learning policy tailored to congestion control in front-haul networks. Experience in the form of state-action pairs is then collected from this baseline policy, and deep symbolic regression is applied to the gathered dataset. This process results in closed-form expressions that approximate the performance of the baseline policy, such as link utilization, delay, and fairness, and these expressions can be directly implemented in any programming language, overcoming limitations during inference. The integration of neuro-symbolic AI offers a novel approach to congestion control in wireless networks, balancing performance, interpretability, and deployment efficiency.
\end{itemize}

\textbf{\emph{2) Solutions for End-to-End Reliability and Latency Guarantee:}} Modern wireless systems must support mission-critical applications such as industrial automation, remote healthcare, and vehicular communications, which demand ultra-high reliability and extremely low latency, making URLLC a cornerstone of 6G services \cite{liu2024bandwidth}. Traditional model-based approaches, while analytically tractable, rely on idealized assumptions that often fail under real-world wireless conditions \cite{she2020deep}. Existing transport protocols, including TCP, Quick UDP Internet Connections (QUIC), and Hybrid Automatic Repeat Request (HARQ), exhibit inherent limitations: TCP prioritizes throughput over latency, QUIC lacks adaptability to dynamic radio environments, and HARQ’s reliance on retransmissions introduces variable delays that hinder reliability guarantees \cite{bennis2018ultrareliable}, \cite{ding2023harq}. Operating within static, rule-based frameworks, these protocols cannot dynamically adapt to bursty traffic, fluctuating channels, or stringent timing constraints, leaving them ill-suited for URLLC scenarios.

To address these shortcomings, recent research has turned to AI reasoning approaches for adaptive reliability and latency control. These approaches encompass methods, such asSelf-Supervised Learning (SSL), which enable responsive and explainable adaptation in URLLC scenarios.

\begin{itemize}
   % \item \textbf{CoT and ToT Reasoning:} In URLLC scenarios, CoT and ToT reasoning allow transport agents to make structured reliability-latency tradeoffs. CoT can analyze factors such as delay bounds and retransmission budgets in a staged manner  \cite{wei2022chain}. ToT enhances this by simulating multiple minislot scheduling options or HARQ strategies in parallel  \cite{yao2023tree}, enabling informed decisions under real-time constraints. These approaches support dynamic, interpretable latency control, especially in time-sensitive industrial and vehicular applications.
    \item \textbf{Self-Supervised Learning:} It has emerged as a pivotal technique in wireless networks to address challenges related to end-to-end reliability and latency guarantees. By leveraging unlabeled traffic data, SSL models can discern subtle patterns in network behavior, enhancing the system's ability to adapt to dynamic conditions without the need for extensive labeled datasets  \cite{achiam2023gpt}. By distinguishing stable from unstable network states, these models improve generalization to unseen channel fluctuations or congestion bursts. This capability is crucial for maintaining predictive accuracy and delay resilience in real-time low-latency applications, such as autonomous driving or industrial automation, where timely and reliable communication is paramount. The ability to learn from unlabeled data eliminates the need for costly annotations and accelerates the deployment of adaptive communication systems, making SSL an efficient solution for dynamic wireless environments.
\end{itemize}

\textbf{\emph{3) Solutions for Semantic Communication:}} Semantic communication has emerged as a key enabler for future wireless networks by transmitting not only raw data but also the underlying intent and contextual meaning, thereby reducing redundancy, improving resource allocation, and supporting intelligent interaction in dynamic environments \cite{luo2022semantic}, \cite{liang2024generative}. As 6G systems approach the limits of conventional modulation, coding, and compression, particularly under strict reliability and latency constraints, semantic communication offers a path to enhanced efficiency and situational awareness. However, existing methods often restrict semantics to predefined label sets, neglecting implicit or unstructured meaning \cite{liang2022life}, while traditional machine learning frameworks struggle with scalability and contextual reasoning in complex, dynamic settings \cite{getu2023tutorial}. Addressing these gaps is critical for building flexible, robust, and reasoning-aware semantic communication systems.

To tackle these limitations, researchers have investigated AI reasoning approaches for semantic communication. These include RAG and KG-based reasoning with GNNs, both of which enhance contextual understanding and adaptability in dynamic wireless environments.

\begin{itemize}
    \item \textbf{RAG Method:} RAG enhances semantic communication by combining parametric and non-parametric memory. This integration allows models to retrieve external knowledge, such as semantic structures or domain-specific ontologies, during both encoding and decoding. In wireless communication, RAG helps optimize transmission strategies by referencing previous encoding methods that maintained object recognizability under similar network conditions. This retrieval mechanism improves contextual relevance and enables adaptive, history-aware transmission, enhancing the reliability and quality of communication under dynamic channel conditions. For instance, during video transmission over unreliable links, RAG’s ability to reference past successful strategies ensures efficient encoding and minimizes packet loss, even in fluctuating environments  \cite{lewis2020retrieval}. By integrating retrieval-based strategies with generation, RAG ensures that transmitted data aligns with user intent and network dynamics, offering a scalable and efficient approach to semantic communication in wireless networks.
    \item \textbf{KG-GNN:} GNNs combined with domain-specific KGs provide a powerful framework for semantic dependency reasoning within transmitted content. When applied at the transport layer, it can model inter-entity relationships such as temporal causality, spatial containment, or functional relevance. For example, in a real-time surveillance video, the GNN can identify that the presence of a person near a door is more semantically important than background motion. The model then prioritizes the transmission of features related to these entities, guided by KG-based reasoning \cite{salehi2025llm}. This selective compression and transmission process preserves the semantic integrity of the content while reducing spectral overhead, and is particularly suited for 6G applications involving multimodal, real-time semantic flows.
\end{itemize}

\subsection{Typical Scenarios}

\textbf{\emph{1) Cross-Scenario Congestion Control:}}
In heterogeneous wireless environments such as 5G, Wi-Fi, and satellite communications, neuro symbolic reasoning has been applied to enable interpretable and generalizable congestion control strategies. By modeling the relationship between key link performance indicators such as RTT, packet loss rate, and queuing delay, and the resulting transmission behavior, deep symbolic regression methods can automatically derive closed-form control expressions for window adjustment \cite{martins2024closed}. These symbolic expressions are directly deployable on network nodes, providing transparent and lightweight decision-making across diverse transport scenarios. The neuro-symbolic framework integrates neural representation learning with symbolic logic, allowing the system to adapt to varied network conditions while preserving interpretability. This approach has demonstrated superior steady-state performance and responsiveness to burst traffic compared with traditional TCP variants and reinforcement learning-based methods. It is particularly well suited for large scale deployments involving edge devices and mobile terminals, where low complexity and adaptive congestion control is essential.

\textbf{\emph{2) Semantic Image Transmission in Urban Macrocell Environments:}}
In 6G urban macrocell scenarios, real-time image transmission is constrained by limited uplink bandwidth and variable channel conditions. A recent semantic communication framework integrates LLMs with neuro-symbolic reasoning to enhance transmission efficiency and robustness \cite{ribouh2025large}. While semantic extraction occurs at the application layer, the transport layer adjusts the redundancy of symbolic representations based on real-time link metrics such as SNR and subcarrier quality. This enables adaptive compression and symbol prioritization tailored to network conditions.
Neuro-symbolic reasoning supports interpretable transport decisions by selecting essential semantic entities and structuring them into compressible symbolic formats, while neural components provide generalization across channel dynamics. The system achieves compression ratios over 4250 to 1, with better semantic reconstruction than traditional methods. This approach highlights how AI reasoning can guide adaptive scheduling and redundancy control in the transport layer, ensuring semantic fidelity and efficiency under bandwidth-constrained wireless environments.

\subsection{Lessons Learned}

The integration of AI reasoning methods, such as CoT, ToT, and RAG, has notably improved transport layer performance in managing congestion control, reliability, and latency. These advancements improve network performance, reduce latency, and enhance cache hit rates. Ongoing research will focus on refining AI systems to address cross-layer interactions and multi-modal data handling, ensuring future systems meet the needs of 6G and beyond.

\section{AI Reasoning for Application Layer}\label{sec-VII}

As the top layer of the communication stack, the application layer connects users and services, serving as the final gateway for delivering network value. Beyond data presentation and user interaction, it plays an active role in content control, quality of experience modeling, multimodal fusion, and task scheduling. In semantically driven architectures, it is becoming a core decision-making entity, guiding cross-layer coordination based on user intent. Its intelligence will increasingly influence network efficiency and user experience, especially across cloud, edge, and device systems.

\begin{figure*}[!t]
\centering
 \includegraphics[width=5.5 in]{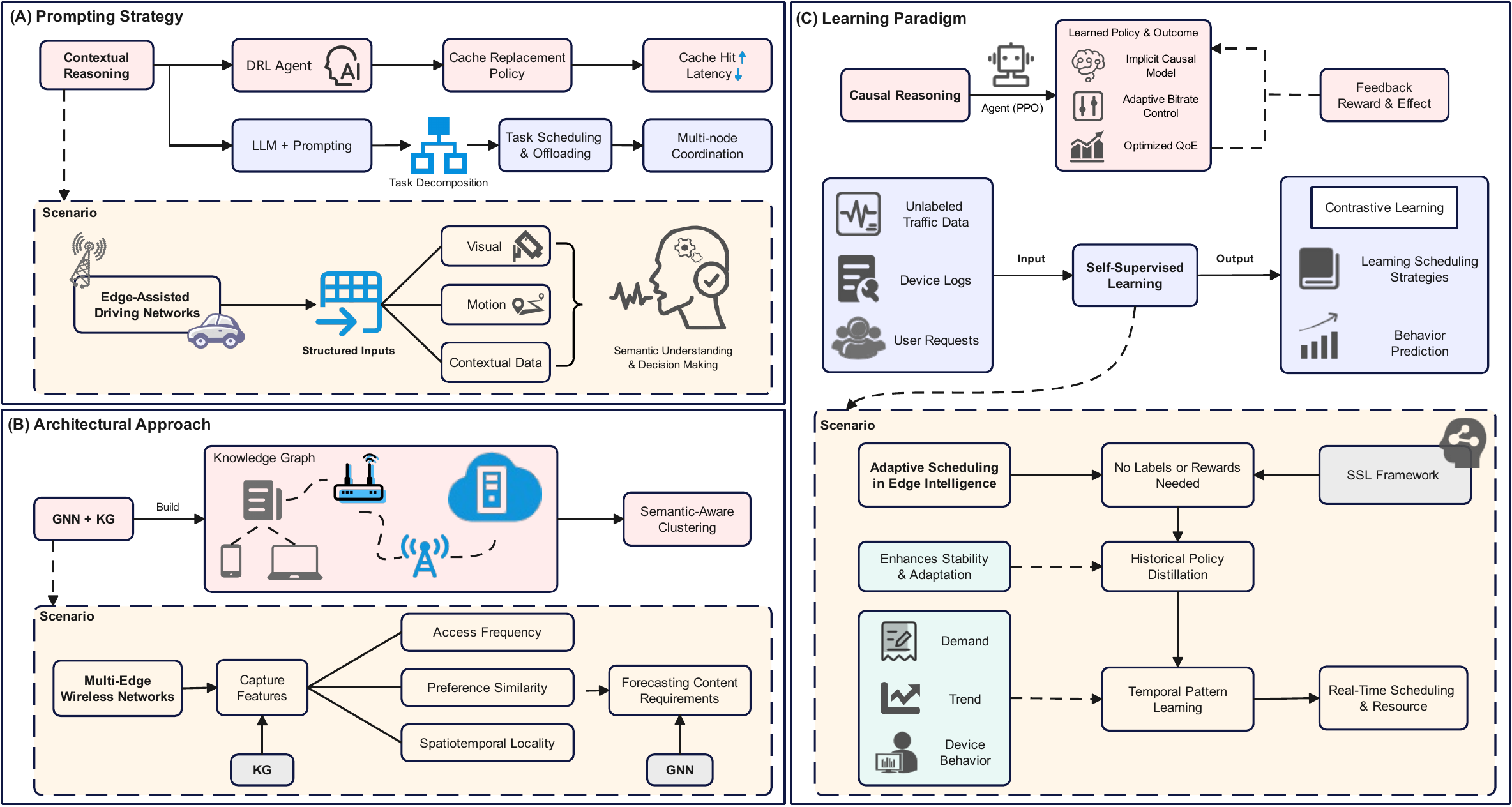}
   \caption{AI reasoning methods for the application layer. (A) Prompting strategies, such as contextual reasoning, support context-aware decision making \cite{zhou2024generative} \cite{huang2024efficient}, enabling semantic reasoning in caching and edge intelligence. (B)Architecturally, KG-GNN captures spatiotemporal and preference relations \cite{wang2024knowledge}, while structured inputs and multimodal fusion support real-time behavior understanding. (C) In learning paradigms, SSL enhances adaptability and scheduling stability in dynamic environments \cite{naresh2023deep}, while causal reasoning truly shapes user experience over time \cite{naresh2023ppo}.}
    \vspace{-0.5cm}
\label{fig: AI Reasoning Method in Application Layer}
\end{figure*}

The application layer exhibits distinct characteristics, such as semantic complexity in functions like compression, summarization, and task scheduling, which require a deep understanding of user behavior, multimodal inputs, and content structure. Additionally, operational heterogeneity arises from managing diverse application types, device capabilities, and service demands. Dynamic adaptability is crucial to adjust to changes in user intent, device status, and network feedback. These challenges make static rule-based systems and shallow learning techniques inadequate \cite{wei2021wireless}. AI reasoning methods, such as LLM agents, neuro-symbolic inference, and self-supervised reinforcement learning, offer the necessary capabilities for intelligent and context-aware application-layer control. Fig. \ref{fig: AI Reasoning Method in Application Layer} illustrates AI reasoning methods for the application layer.

\subsection{AI Reasoning Solutions}

\textbf{\emph{1) Content Distribution and Caching:}}
With the increasing demand for bandwidth-intensive services like video streaming, AR/VR, and immersive communications, modern wireless networks face challenges in content availability, speed, and reliability. Edge caching has become a key solution to reduce dependence on centralized servers and alleviate backhaul congestion, ensuring low-latency access near users \cite{wang2024knowledge}. However, traditional methods such as heuristic and optimization-based strategies struggle with scalability or high computational costs, limiting their real-time use \cite{wang2024knowledge}. Fixed-rule caching policies, like least-recently used (LRU) and least-frequently used (LFU), also fail to adapt to dynamic user behavior and spatiotemporal content shifts, and lack semantic awareness, making it difficult to account for popularity trends, user intent, and contextual relevance \cite{wei2021wireless}, \cite{li2024scalm}. To address these challenges, AI-driven approaches incorporating cognitive reasoning have been introduced, enabling semantic-aware scheduling, mobility-aware content migration, and collaborative prefetching based on learned patterns \cite{li2020cooperative}, \cite{ammanabrolu2018playing}.

To overcome these limitations, recent research has introduced AI-driven approaches that incorporate cognitive reasoning into edge caching. Such methods enable semantic-aware scheduling, mobility-aware content migration, and collaborative prefetching based on learned patterns. These solutions leverage advanced techniques are as followed.

\begin{itemize}
    \item \textbf{KG-GNN:} A reasoning approach based on GNN and KGs has been proposed to enhance cooperative caching in MEC-enabled wireless systems. The system constructs a KG that captures semantic relationships among users, content, and edge locations, incorporating features such as access frequency, preference similarity, and spatiotemporal locality \cite{wang2024knowledge}. GNN operates on this structured representation to infer latent user-content affinities and predict future content demand across edge regions. By embedding user behavior and content metadata into graph nodes, the model enables semantic-aware content clustering, proactive replication, and collaborative cache prefetching. This graph-based reasoning framework supports multi-hop dependency learning and improves caching coordination, ultimately enhancing content availability and reducing latency in dynamic and heterogeneous wireless environments.
    \item \textbf{Contextual Reasoning:} 
    Adaptive Contextual Caching (ACC) method is introduced as a solution to optimize content caching in mobile-edge LLMs, addressing the limitations of traditional caching approaches. In mobile-edge LLM services, ACC utilizes DRL to dynamically optimize cache replacement policies \cite{liu2025adaptive}. By considering factors such as semantic similarity, user behavior, and the frequency of content updates, the ACC framework proactively determines what to cache and when to replace cached content. For example, in scenarios like autonomous driving, where the edge LLM needs to respond to real-time queries, ACC predicts and stores relevant traffic or legal data based on previous user interactions and contextual information, reducing retrieval latency and computational overhead. Overall, ACC significantly enhances the efficiency of content distribution and caching in wireless networks, making it an essential tool for addressing the dynamic and high-demand nature of modern communication systems.
\end{itemize}

\textbf{\emph{2) Distributed Applications Collaboration:}}
In scenarios such as edge intelligence, industrial automation, and the IoT, distributed devices must collaboratively handle tasks like real-time sensing, multimodal fusion, and task offloading. These tasks are often heterogeneous and dynamic, with varying user preferences and service demands. Without intelligent coordination, systems face degraded performance and inefficient resource use \cite{zhou2024generative}. Traditional heuristic-based scheduling methods emphasize local optimization but lack global perspective, limiting their ability to support multi-device collaboration under dynamic workloads \cite{asghar2022survey}. They often ignore user intent, context, and semantic dependencies, resulting in poor generalization and suboptimal resource allocation. These challenges underscore the need for AI reasoning frameworks that incorporate logic inference, contextual awareness, and task decomposition to support scalable and adaptive scheduling in distributed wireless systems.

To address these challenges, recent research has proposed several AI reasoning methods for distributed applications collaboration. These methods include contextual reasoning, both of which enhance coordination efficiency and task allocation across distributed devices.

\begin{itemize}
    \item \textbf{Contextual Reasoning:} In-context generation reasoning enables LLM based agents to adapt to dynamic task offloading scenarios without requiring parameter fine tuning. Rather than learning fixed policies, the model receives context rich prompts that encode task attributes, service demands, network load, and device capabilities, then generates scheduling or placement decisions accordingly \cite{zhou2024generative}. This method belongs to the prompting strategy category and leverages the LLM’s latent reasoning ability to decompose high level task specifications into executable actions. It is particularly suitable for edge cloud environments where workload patterns and service types are highly variable. By interpreting prompts that resemble historical offloading traces, the model can dynamically align its decisions with real time system conditions. In the context of distributed applications collaboration, this reasoning paradigm enhances coordination efficiency and responsiveness by enabling prompt based, context aware decision making across multiple edge and cloud nodes.
%    \item \textbf{GNN:} A graph-structured reinforcement reasoning approach has been proposed to support intelligent task scheduling in distributed edge computing systems. Tasks, devices, and network resources are modeled as nodes in a heterogeneous graph that captures semantic relationships such as task dependencies, resource availability, and communication costs. A GNN encodes this structure to extract contextual embeddings, while a MARL policy enables decentralized agents to collaboratively determine task placement and resource allocation \cite{li2023task}. Each agent uses local graph features to reason about its environment and optimize decisions based on long-term goals such as reducing task delay and improving execution reliability. This method integrates spatial constraint reasoning with system dynamics, enhancing coordination efficiency in complex edge clusters. It demonstrates superior task completion and resource utilization across dynamic scheduling scenarios, validating the scalability and adaptability of GNN-RL reasoning in wireless edge intelligence systems.
\end{itemize}

\textbf{\emph{3) Appropriate Perception of QoE:}}
Application-layer services like video conferencing, online gaming, and AR/VR are highly sensitive to user-perceived QoE, which is influenced by factors such as delay, bandwidth variation, compression quality, and scene complexity \cite{ajeyprasaath2024hybrid}. However, QoE modeling is challenging due to the variability of user satisfaction across different contexts. Traditional methods, including heuristics and regression, assume simple linear relationships and fail to capture the complex interactions among visual, auditory, and behavioral signals. These models struggle with the high-dimensional, dynamic nature of QoE. Furthermore, conventional scheduling policies like Buffer Based (BB), Buffer Occupancy-based Lyapunov Algorithm (BOLA), and Rate-based (RB) often lack semantic awareness, leading to poor adaptability and inaccurate predictions \cite{naresh2023deep}, \cite{naresh2023ppo}. These issues highlight the need for advanced reasoning methods that can interpret multimodal semantics and user preferences in a context-aware manner.

To address these challenges, recent approaches leverage AI reasoning frameworks to enhance QoE perception, with causal reasoning offering promising solutions.

\begin{itemize}
    \item \textbf{Causal Reasoning:} In dynamic control systems, causal reasoning involves learning the consequences of interventions to steer the system towards a desired state. This principle is powerfully demonstrated in adaptive bitrate streaming. Here, an agent implementing the PPO method must reason about how its decisions will causally affect future Quality of Experience or QoE \cite{naresh2023ppo}. The agent performs a series of interventions by selecting a bitrate based on observed conditions like network bandwidth and playback buffer. The critical step is learning the cause and effect relationship by understanding how a specific bitrate choice impacts subsequent QoE metrics like video smoothness, quality variation, and rebuffering duration. This causal knowledge is acquired through active experimentation and trial and error, guided by a reward function that defines the desirable outcome. Its key advantage in addressing the Appropriate Perception of QoE challenge is that the resulting policy moves beyond simple metric correlation to embody a deep, context aware understanding of how control actions truly shape user experience over time.
\end{itemize}

\subsection{Typical Scenarios}

\textbf{\emph{1) Multi-Edge Caching Networks:}}
In MEC enabled heterogeneous wireless networks, collaborative caching among edge nodes plays a critical role in improving content availability and reducing access latency. This is especially important in scenarios such as video-on-demand and real-time news dissemination, where content popularity shifts dynamically and user requests are spatially distributed. A recent approach combines KG-based semantic reasoning with deep reinforcement learning to support intelligent caching coordination \cite{wang2024knowledge}. The system constructs a KG representing relationships among users, content, and edge nodes, encoding factors such as access frequency, user preferences, and spatial locality. Based on this structured representation, a Dueling DQN agent learns content migration and replacement strategies by interacting with the graph over time. This reasoning framework enables edge nodes to share caching states, reason over content semantics, and jointly optimize storage usage, highlighting the effectiveness of graph-based and learning-driven reasoning in application-layer edge intelligence.

\textbf{\emph{2) Edge-Assisted Driving Networks:}}
In application layer scenarios involving intelligent driving and vehicular services, systems must interpret real-time behavior and environmental dynamics beyond data transmission. A reasoning framework based on LLMs with multimodal prompting has been proposed to support edge-side behavior understanding and semantic decision-making. This approach deploys LLMs on roadside edge units (RSUs), where models receive structured inputs combining visual, motion, and contextual data from cameras and radar sensors        \cite{huang2024efficient}. Using prompts that encode environment, objects, and actions, the LLM performs real-time reasoning to narrate driving behaviors. This enables a closed-loop system where inference and interaction occur locally, reducing reliance on cloud resources and improving latency. The method has proven effective in autonomous driving, traffic monitoring, and assisted decision making, where low latency and interpretability are essential. Contextual prompting with LLM reasoning enhances adaptability to diverse traffic contexts and supports semantic services at the edge.

%\textbf{\emph{3) Adaptive Scheduling in Edge Intelligence:}}
%In application-layer wireless systems involving content delivery, computation offloading, and caching, scheduling decisions must be made in real time under dynamic network conditions and limited feedback. Self-supervised reinforcement learning (SSL) provides a reasoning framework that enables agents to derive optimal strategies through environment interaction and experience distillation. A recent SSL-based system integrates historical policy distillation to enhance stability and accelerate policy adaptation without requiring manual labels or predefined rewards \cite{naresh2023deep}. By combining auxiliary self-supervised objectives with reinforcement learning, the system captures temporal patterns in user demand, content trends, and device behavior to guide scheduling and resource allocation. This approach suits distributed edge networks where agents coordinate resources across heterogeneous nodes and shifting workloads. SSL-based methods outperform conventional RL baselines in sample efficiency, robustness, and accuracy, offering a scalable and adaptive solution for application-layer intelligence in wireless systems.

\subsection{Lessons Learned}
The integration of AI reasoning methods has greatly enhanced the application layer's ability to manage QoS/QoE, content distribution, and adaptive scheduling. These approaches enable intelligent, real-time decision-making in dynamic environments, overcoming traditional method limitations. AI systems improve resource allocation and content delivery by leveraging semantic reasoning and contextual analysis, making them particularly effective in latency-sensitive applications like AR/VR. The ability to adapt to changing network conditions and user behaviors ensures that the application layer can meet performance and reliability demands. The continued evolution of AI reasoning will be key to addressing the growing complexities of 6G.

\section{Future Research Directions} \label{sec-X}

\subsection{Agentic AI Reasoning}

Future networks will become increasingly complex and dynamic, with massive devices and stringent service demands. This drives the need for autonomous LLM-based agents that can intelligently plan and execute real-time network management tasks without constant human oversight \cite{zhang2025toward}. 
Using advanced reasoning, such agents could interpret high-level goals and devise multi-step strategies, such as reconfiguring parameters or scheduling resources, to achieve them. Approaches like CoT and ReAct combine logical reasoning with action, letting LLM ``network brains” break down complex tasks and interact with environments step-by-step. It enables more adaptive, proactive network orchestration, where AI agents monitor conditions, predict issues, and coordinate responses.

Today’s LLM agents are still immature for telecom operations. Off-the-shelf LLMs lack precise domain knowledge of wireless protocols and constraints, risking hallucinated or unsafe actions if deployed directly. Additionally, purely text-based agents struggle with real-time data integration and can’t yet directly execute network controls. 
Key directions include grounding LLM agents in wireless environments by linking them to network state databases, simulators, and APIs, enabling safe reasoning with live data and action execution. Moreover, multiple collaborative agents could specialize, e.g., one managing radio resources, another handling core networks, and coordinate via natural language to tackle complex multi-domain scenarios \cite{luo2025toward}. Ultimately, a reliable agentic reasoning paradigm could bring about self-driving networks that continuously learn and adapt strategies for optimal performance.

\subsection{Data-Efficient Learning and Generalization}
A key challenge for applying AI reasoning in wireless networks is the scarcity of domain-specific training data \cite{sun2025comprehensive}. Wireless environments are diverse and fast-evolving, making large labeled datasets for each scenario impractical. Thus, data-efficient learning is crucial. LLMs offer advantages like in-context learning, adapting to new tasks from a few examples without extra training, valuable in wireless settings, where LLMs could generalize from demonstrations. 

Despite broad knowledge, base LLMs struggle with wireless-specific tasks, as telecom details, including protocols, radio physics, are underrepresented in their training. Generic LLMs may perform poorly on domain inputs. Fine-tuning on wireless data helps but requires costly niche datasets, and models tuned for one domain, e.g., 5G, often fail to generalize to others. Promising solutions include cross-domain adaptive models using meta-learning or few-shot transfer to adapt between contexts \cite{ding2024load}. Meanwhile, creating shared communication representations, pretrained on logs, manuals, and simulations, can also provide a strong foundation for downstream tasks. Combining few-shot prompting, lightweight fine-tuning, and cross-domain transfer can enable robust generalization from minimal real-world wireless data.
\subsection{Efficient Deployment and Collaboration of Large Models}
State-of-the-art reasoning models with tens of billions of parameters are resource-intensive, posing challenges for wireless edge deployment. In networks like 6G, many intelligent functions need to run at the edge or on user devices for low latency and privacy, but these environments have limited computation, energy, and bandwidth. Naively deploying large LLMs in each base station or handset is unfeasible. Thus, researchers are exploring splitting and specializing reasoning pipelines to leverage large models under real-world constraints. One approach is workload partitioning. Edge devices or base stations execute initial layers or simple reasoning locally, offloading complex processing to cloud/core servers. This reduces latency for early responses and confines heavy computation to infrastructure \cite{priya2024energy}. 

However, seamless collaboration is challenging. Splitting reasoning risks communication delays, inefficient data transfer, and lost reasoning coherence, with no standardized protocols. Even trimmed-edge LLMs strain resources. Also, quantization helps, but extreme compression may harm reasoning.  Promising innovations include hierarchical reasoning \cite{yang2025reasonflux}, on-demand model loading \cite{yang2024demand}, and collaborative frameworks \cite{ferrag2025llm}. Such approaches align with network virtualization trends, enabling powerful AI across wireless networks without overburdening edge resources.

\subsection{Safe and Reliable AI Reasoning}
Reliability is critical in wireless communications, as network control errors can cause outages, quality issues, or safety risks, e.g., in connected vehicles \cite{xu2022wireless}. Thus, AI reasoning systems here must be trustworthy, transparent, and robust. Unlike traditional algorithms, LLM-based reasoning may generate hallucinations or cascading logical errors in multi-step processes \cite{huang2025survey}. Ensuring accuracy and uncertainty flagging is key for operator trust.

    Current LLMs lack guarantees of correctness or alignment with domain constraints, risking invalid solutions. They are vulnerable to adversarial inputs, and their opaque decision-making hinders verification. Few mechanisms exist to detect reasoning failures pre-deployment, forcing manual verification that undermines automation efficiency. To build safe systems, researchers are exploring feedback loops and self-verification \cite{stechly2024self}. LLMs or a second model can critique their reasoning, like proof checkers, with multi-path reasoning, improving accuracy. Meanwhile, domain validators, e.g., network simulators and rule engines \cite{ma2024klonet}, can test LLM-proposed actions against constraints, using tool-augmented frameworks like ReAct. Self-refinement and human oversight with wireless-specific rationales, confidence scores also help. Combining rigorous verification, self-checking, and transparency will enable reliable autonomous network control.
    
\section{Conclusion} \label{sec-XI}
This survey comprehensively reviewed the emerging role of AI reasoning in next-generation wireless communications and networks. We systematically categorized reasoning-aware AI techniques into prompting, architectural, and learning paradigms, and examined their layered applications from the physical layer up to security and cross-layer orchestration. The survey reveals that traditional black-box AI struggles with dynamic environments, interpretability, and generalization. In contrast, reasoning-based AI offers transparent, step-by-step decision making, seamless integration of domain knowledge, and robust adaptation to unseen conditions. Representative case studies across UAV networks, massive MIMO, IoT random access, URLLC transport slicing, and zero-day threat detection demonstrate significant performance gains, reduced human intervention, and improved trust. Looking forward, four key research directions are identified, including agentic AI reasoning,  data-efficient learning and generation, efficient deployment and collaboration of LLMs, and safe, verifiable, and robust reasoning frameworks. By bridging AI and communications perspectives, this survey charts a practical path toward cognitive, self-driving future networks that continuously reason, learn, and optimize in real time.

\vspace{-0.2cm}
\bibliographystyle{ACM-Reference-Format}
\bibliography{sample-base}

% \appendix

\end{document}